\documentclass[11pt,a4paper]{article}

\usepackage{jheppub}

\RequirePackage{ifpdf} 
\usepackage{amsmath} 
\usepackage{mathtools}
\usepackage{pstricks}
\usepackage[final]{pdfpages} 
\usepackage{ifpdf} 
\usepackage{slashed}
\usepackage{hyperref}
\usepackage{color} 
\usepackage{graphics}
\usepackage{lineno}

\usepackage{etoolbox} 
\usepackage{fixmath}
\usepackage{amsfonts}

\usepackage{multirow}
\usepackage{epstopdf}

\allowdisplaybreaks[4]

\title{\boldmath Three loop form factors of a massive spin-2 particle with nonuniversal 
coupling}
\author{Taushif Ahmed$^{a,b,}$\footnote{Also at Institute for Theoretical Particle Physics, KIT, 76128, Karlsruhe, Germany from 1st October 2016.}, Pulak Banerjee$^{a,b}$, Prasanna K.\
  Dhani$^{a,b}$, Prakash Mathews$^c$, Narayan Rana$^{a,b}$ and V. Ravindran$^{a}$} 
  
  \affiliation{$^a$ The Institute of Mathematical Sciences, Taramani,
  Chennai 600113, India \\ $^{b}$ Homi Bhaba National Institute,
  Training School Complex, Anushakti Nagar, Mumbai 400085, India \\
  $^{c}$Saha Institute of Nuclear Physics, 1/AF Bidhan Nagar, Kolkata 700 064, West Bengal, India}

\emailAdd{taushif@imsc.res.in}
\emailAdd{bpulak@imsc.res.in}
\emailAdd{prasannakd@imsc.res.in}
\emailAdd{prakash.mathews@saha.ac.in}
\emailAdd{rana@imsc.res.in}
\emailAdd{ravindra@imsc.res.in}

\abstract{
We investigate the interaction of spin-2 fields with those of the Standard
Model in a model independent framework.  We have considered interactions
where the spin-2 fields couple to two sets of gauge invariant tensorial
operators that are not conserved unlike the energy momentum tensor with
different coupling strengths.  Such interactions not only change the
ultraviolet behaviour of the couplings but also expand the scope of the
searches of spin-2 particles at the colliders.  We present all the
relevant renormalisation constants up to three loop level in QCD and also
the form factors that contribute to potential observables.  This sets
the ground to investigate the phenomenological consequences of these
interactions with spin-2 fields through more than one tensorial operator.     
}

\begin{document} 

\maketitle




\def\D{{\cal D}}
\def\DD{\overline{\cal D}}
\def\g{\overline{\cal G}}
\def\gm{\gamma}
\def\M{{\cal M}}
\def\uM{{\hat{\cal M}}}
\def\ep{\epsilon}
\def\epm1{\frac{1}{\epsilon}}
\def\epm2{\frac{1}{\epsilon^{2}}}
\def\epm3{\frac{1}{\epsilon^{3}}}
\def\epm4{\frac{1}{\epsilon^{4}}}
\def\unM{\hat{\cal M}}
\def\ashat{\hat{a}_{s}}
\def\asmur{a_{s}^{2}(\mu_{R}^{2})}
\def\sigbar{{{\overline {\sigma}}}\left(a_{s}(\mu_{R}^{2}), L\left(\mu_{R}^{2}, m_{H}^{2}\right)\right)}
\def\sigbarn{{{{\overline \sigma}}_{n}\left(a_{s}(\mu_{R}^{2}) L\left(\mu_{R}^{2}, m_{H}^{2}\right)\right)}}
\def\unas{ \left( \frac{\hat{a}_s}{\mu_0^{\epsilon}} S_{\epsilon} \right) }
\def\rnM{{\cal M}}
\def\bt{\beta}
\def\cD{{\cal D}}
\def\cC{{\cal C}}
\def\ca{\text{\tiny C}_\text{\tiny A}}
\def\cf{\text{\tiny C}_\text{\tiny F}}
\def\ct{{\red []}}
\def\sv{\text{SV}}
\def\murOmu{\left( \frac{\mu_{R}^{2}}{\mu^{2}} \right)}
\def\bb{b{\bar{b}}}
\def\bt0{\beta_{0}}
\def\bt1{\beta_{1}}
\def\bt2{\beta_{2}}
\def\bt3{\beta_{3}}
\def\gm0{\gamma_{0}}
\def\gm1{\gamma_{1}}
\def\gm2{\gamma_{2}}
\def\gm3{\gamma_{3}}
\def\nn{\nonumber}
\def\l{\left}
\def\r{\right}

\newcommand{\dis}{}
\newcommand{\overbar}[1]{mkern-1.5mu\overline{\mkern-1.5mu#1\mkern-1.5mu}\mkern
1.5mu}

\section{Introduction}
The discovery of the Higgs boson by ATLAS \cite{Aad:2012tfa} and 
CMS \cite{Chatrchyan:2012xdj} collaborations at
the LHC has put the Standard Model (SM) on a strong footing. The ongoing
precise measurements with the Higgs boson will shed light on the nature
of its coupling to the particles of the SM.  While the SM has been
enormously successful, it is not a complete theory of particle physics.  For
example, it does not accommodate dark matter, non-zero neutrino mass
etc.  These are some of the compelling reasons to go further to
investigate the physics beyond the SM (BSM).  In this context, supersymmetric
extensions of the SM have been studied intensively both theoretically
as well as experimentally.  Similarly, models with extra dimensions,
the strong contenders to SUSY, have also been studied extensively.
These models contain extra degrees of freedom through additional spin-2
bosons.  

In the ADD \cite{ArkaniHamed:1998rs,Antoniadis:1998ig,ArkaniHamed:1998nn} and the RS \cite{Randall:1999ee} models spin-2 particles couple to the SM particles
through energy momentum (EM) tensor of the SM with a single coupling
denoted by $\kappa$.  The phenomenology with this universal
coupling has been studied rigorously.  There are also phenomenological
investigations with spin-2 particles with non-universal coupling
to the SM particles.  While the later may not belong to any of the popular
extra-dimensional models, they can provide an opportunity to study
the distinct signatures at the colliders which is not possible with
theories with universal couplings.  Independent of the nature of
these couplings, these are effective theories and hence
non-renormalizable in the conventional sense.  In the ADD and the
RS, thanks to conservation of EM tensor of the SM, the
leading interaction term that describes the coupling of spin-2
with those of the SM does not require any additional renormalisation.
To this order in the coupling, which is good enough for all the
phenomenological studies, the infrared (IR) structure of the SM is also
not affected and hence factorisation properties continue to hold.
This allows us to compute successfully various observables beyond
leading order in the SM coupling using perturbative methods.  All the
infrared singularities do cancel giving finite perturbative results
that can be used to constrain the model parameters unambiguously.
While this is true with theories with interaction term containing
conserved EM tensor, it is not clear how the ultraviolet (UV) and
IR structure would look like when spin-2 couples to particles of
the SM with different (non-universal) couplings.

Soon after the discovery of the 125 GeV boson, these models with 
spin-2 non-universal coupling, have become important in the context
of imposter to the Higgs boson \cite{Ellis:2012jv, Ellis:2012mj,
Ellis:2013ywa}.  
This was necessary to extend the scope of spin-2 models as the 
experimental bounds on RS resonance (universal coupling) was
much higher \cite{Abazov:2005pi,Aaltonen:2010cf,Aad:2011mh,Aaltonen:2011xp,Chatrchyan:2011wq,Chatrchyan:2011fq,ATLAS:2011ab,Aad:2012cy}.
To NLO in QCD the UV and IR behaviour for the
non-universal couplings for a spin-2, had been studied in the
context of Higgs Characterisation \cite{Artoisenet:2013puc}.  UV
renormalisation is needed as a result of the non-universal
couplings and with regard to the IR structure, the double and
single pole terms contained the appropriate  universal IR
coefficients that canceled against those coming from real emission
processes and mass factorisation counterterms.  This was a
demonstration of IR factorisation to NLO for the non-universal
couplings scenario \cite{Artoisenet:2013puc}.  Recently, in the
context of the 750 GeV spin-2 resonance with non-universal
couplings have been considered \cite{Das:2016pbk} to NLO+Parton Shower (PS)
accuracy.

The SM at the LHC is being scrutinised at an unprecedented level of 
precision.  It is only natural to have the competing BSM scenarios,
match the same order of accuracy in QCD as the SM
observables.  At hadron collider, the first step to such a 
phenomenological study would be to compute form factors to 
the production of a singlet on shell state $X$ {\em via} the
quark $q \bar q \to X$ or gluon $g g \to X$ production channels,
to the same order of accuracy.  Presently, form factors are
available to up to three-loop level in
the SM \cite{Moch:2005tm, Moch:2005id, Baikov:2009bg,
Gehrmann:2010ue, 
Gehrmann:2014vha}, for some
BSM spin-2 that couples to the
EM tensor \cite{deFlorian:2013sza, Ahmed:2015qia}
and for the pseudo-scalar Higgs boson \cite{Ahmed:2015qpa}.
NLO QCD corrections have been computed for the extra dimension models
{\em viz.}\ ADD and RS for most of the di-final state process 
\cite{Mathews:2004xp,
Mathews:2005zs, Kumar:2006id, Kumar:2008pk, Kumar:2009nn, Agarwal:2009xr,
Agarwal:2010sp, Agarwal:2010sn} and 
this has been extended to NLO+PS accuracy \cite{Frederix:2012dp,
Frederix:2013lga, Das:2014tva}.   Recently, for the di-lepton production
to NNLO order in QCD for the ADD model was computed 
\cite{Ahmed:2016qhu}.

In this article, we investigate the UV structure of the interaction
term up to three loop level in QCD.  We restrict ourself to QCD
sector of the SM because the phenomenology with such operators have
immediate application at the LHC where such interactions are probed
by strong interaction.  There are of course many ways spin-2 can
couple to the SM.  Here, we study the minimal version where spin-2 fields
couple to QCD through two different operators with two different
couplings, each operator is invariant under gauge group of QCD.  Note
that spin-2 is gauge singlet.  These operators are rank-2 but
unfortunately not conserved unlike EM tensor of QCD~\cite{Nielsen:1977sy}.  As
a consequence of this, both the operators as well as the couplings get
additional UV renormalisation order by order in perturbation theory.

In addition, we intend to compute the on-shell form factors of these
operators between quark and gluon states that are important ingredients
of any observable at the LHC to study such interactions.  In this
article we achieve this by computing the on-shell form factors of
these operators.  This is possible, thanks to the universal IR structure
QCD amplitudes even in the case of a non-universal spin-2 coupling.

In section 2, we describe the theoretical framework, which include 
the details of the interaction Lagrangian, UV and IR renormalisation
procedure.  Computational details, the unrenormalized form factors
and anomalous dimensions to three loop level are given in section 3.
Finally, we present our conclusions in section 4. 

\section{Theoretical Framework}
\subsection{The Effective Action}
The minimal effective action that describes the coupling of spin-2 fields
denoted by $h_{\mu\nu}$ with those of QCD consists of two gauge invariant
operators $\hat {\cal O}^{G,\mu\nu}$ and $\hat {\cal O}^{Q,\mu\nu}$ \footnote{This is
not the unique decomposition of original EM tensor.  One can adjust
gauge invariant terms between these two.}:
\begin{equation}
\label{eq:action}
S =\int d^4x ~ {\cal L}_{QCD}-{1 \over 2} \int d^4x ~ h_{\mu\nu}(x) \left(\hat \kappa_G~ \hat {\cal O}^{G,\mu\nu} (x) + \hat \kappa_Q~\hat {\cal O}^{Q,\mu\nu}(x) \right)
\end{equation}
where $\hat \kappa_I, I=G,Q$ are dimensionful couplings.  $G$ denotes the
pure gauge sector, while $Q$ denotes the fermionic sector and its gauge
interaction.  The gauge invariant operators $\hat {\cal O}^{G,\mu\nu}$ and
$\hat {\cal O}^{Q,\mu\nu}$ are given by 
\begin{eqnarray} \label{eq:emtensor}
\hat {\cal O}^G_{\mu\nu} &=&{1 \over 4} g_{\mu\nu} \hat F_{\alpha \beta}^a \hat F^{a\alpha\beta} 
- \hat F_{\mu\rho}^a \hat F^{a\rho}_\nu
 - \frac{1}{\hat \xi} g_{\mu\nu} \partial^\rho(\hat A_\rho^a\partial^\sigma \hat A_\sigma^a)
-{1 \over 2\hat \xi}g_{\mu\nu} \partial_\alpha \hat A^{a\alpha} \partial_\beta \hat A^{a\beta} 
 \nonumber\\
&& + \frac{1}{\hat \xi}(\hat A_\nu^a \partial_\mu(\partial^\sigma \hat A_\sigma^a) + \hat A_\mu^a\partial_\nu
 (\partial^\sigma \hat A_\sigma^a))
+\partial_\mu \overline {\hat \omega^a} (\partial_\nu \hat \omega^a - \hat g_s f^{abc} \hat A_\nu^c \hat \omega^b)
\nonumber\\
&& +\partial_\nu \overline {\hat \omega^a} (\partial_\mu \hat \omega^a- \hat g_s f^{abc} \hat A_\mu^c \hat \omega^b)
-g_{\mu\nu} \partial_\alpha \overline {\hat \omega^a} (\partial^\alpha \hat \omega^a - \hat g_s f^{abc} \hat A^{c \alpha} \hat \omega^b),
\\
 \hat {\cal O}^Q_{\mu\nu} &= &
 \frac{i}{4} \Big[ \overline {\hat \psi} \gamma_\mu (\overrightarrow{\partial}_\nu -i \hat g_s T^a \hat A^a_\nu)\hat \psi
 -\overline {\hat \psi} (\overleftarrow{\partial}_\nu + i \hat g_s T^a \hat A^a_\nu) \gamma_\mu \hat \psi
 +\overline {\hat \psi} \gamma_\nu (\overrightarrow{\partial}_\mu -i \hat g_s T^a \hat A^a_\mu)\hat \psi
 \nonumber\\
 &&-\overline {\hat \psi} (\overleftarrow{\partial}_\mu + i \hat g_s T^a \hat A^a_\mu) \gamma_\nu \hat \psi\Big]
- ig_{\mu\nu} \overline {\hat \psi} \gamma^\alpha (\overrightarrow{\partial}_\alpha -i \hat g_s T^a \hat A^a_\alpha)\hat \psi
 \end{eqnarray}
where $\hat A^a_{\mu}$, $\hat \psi$, $\hat \omega^a$ and $h_{\mu\nu}$ are gauge, quark, ghost and spin-2 fields, respectively. 
$\hat g_{s}$ is the strong coupling constant and $\hat \xi$ is the gauge fixing
 parameter.  The hat on all the quantities indicate that they are bare/unrenormalized.  
$T^{a}$ and $f^{abc}$ are the Gell-Mann matrices and structure constants of SU(N)
 gauge theory, respectively.  In the above, we have retained terms only up to order $\hat \kappa$ and
in the rest of the paper, we restrict ourselves to this approximation.   

\subsection{Ultraviolet renormalization}
Note that the sum $\hat {\cal O}^{G,\mu \nu} + \hat {\cal O}^{Q,\mu \nu}$ is nothing but
the EM tensor of QCD. Unlike the EM tensor, neither of these composite operators is individually 
conserved and hence are not protected by QCD radiative corrections.  In other words, they develop additional UV divergences
which need to be factored out in terms of  UV renormalisation constants and then removed by renormalisation procedure. 
This is achieved by renormalizing bare coupling constants $\hat \kappa_I, I=G,Q$ with the help of those renormalisation constants.
The resulting interaction terms expressed in terms of  renormalised operators with appropriate renormalised couplings
are guaranteed to predict UV finite correlation functions to all orders in strong coupling constant. Note that, the operator  $\hat {\cal O}^{G,\mu \nu}$ is free from quark
fields which means in the theory where spin-2 field couples exclusively to the pure Yang-Mills, the operator 
$\hat {\cal O}^{G,\mu \nu}$ is conserved. However, in the presence of the quark fields in QCD, this property ceases to hold true beyond tree level.

The most commonly used method of obtaining the renormalisation constants in quantum field theory 
is to compute off-shell amplitudes and extract the UV divergent contributions order by order in
perturbation theory.  For composite operators, there exists an alternative approach, namely, method 
of the operator product expansion.  We will not follow any of these approaches 
in this article.  Instead, we apply the method
discussed in \cite{Ahmed:2015qpa} to obtain both UV renormalisation constants as well as on-shell form factors of
these operators. In \cite{Ahmed:2015qpa} we have demonstrated that UV renormalisation constants
of composite operators can be extracted order by order in perturbation theory from their
on-shell form factors exploiting their universal IR structure. 
Note that the renormalised on-shell form factors are important components 
of higher order radiative corrections to observables as they give contributions to the pure virtual
part and hence will be useful for further studies. 

We use dimensional regularisation to regulate both UV and IR divergences.  The space-time dimension
is taken to be $d=4+\epsilon$.  Both these divergences appear as poles in $\epsilon$.
Introducing the scale $\mu$ to keep the bare strong
coupling constant ${\hat a}_{s} \equiv {\hat g}_{s}^{2}/16\pi^{2}$ dimensionless,
we relate bare strong coupling constant ${\hat a}_{s}$ to the renormalised one $a_{s} \equiv a_{s} \left( \mu_{R}^{2} \right)$,
at renormalisation scale $\mu_R$, through
\begin{align} \label{eq:asAasc}
  {\hat a}_{s} S_{\epsilon} = \left( \frac{\mu^{2}}{\mu_{R}^{2}}  \right)^{\epsilon/2}
  Z_{a_{s}} a_{s}
\end{align}
with
$S_{\epsilon} = {\rm exp} \left[ (\gamma_{E} - \ln 4\pi)\epsilon/2
\right]$, where $\gamma_{E}$ is the Euler constant.
The renormalization constant $Z_{a_{s}}$ up to
${\cal O}(a_{s}^{3})$ is given by
\begin{align}
  \label{eq:Zas}
  Z_{a_{s}}&= 1+ a_s\left[\frac{2}{\epsilon} \beta_0\right]
             + a_s^2 \left[\frac{4}{\epsilon^2 } \beta_0^2
             + \frac{1}{\epsilon}  \beta_1 \right]
             + a_s^3 \left[\frac{8}{ \epsilon^3} \beta_0^3
             +\frac{14}{3 \epsilon^2}  \beta_0 \beta_1 +  \frac{2}{3
             \epsilon}   \beta_2 \right]\,.
\end{align}
$\beta_{i}$'s are the coefficients of QCD $\beta$ function~\cite{Tarasov:1980au}.

According to the Joglekar and Lee theorem \cite{Joglekar:1975nu} , the two operators ${\cal O}^I$ are closed under renormalization which can be accomplished through
the renormalization mixing matrix $Z$, as follows
\begin{equation}
  \label{eq:Zmat}
\begin{bmatrix}
O^{G} \\ O^{Q}
\end{bmatrix}
=
\begin{bmatrix}
  Z_{GG} & Z_{GQ} \\ Z_{QG} & Z_{QQ}
\end{bmatrix}
\begin{bmatrix}
\hat  O^{G} \\ \hat O^{Q}
\end{bmatrix}
\,.
\end{equation}
The renormalization constants $Z_{IJ}$ satisfy following renormalization group equation (RGE) 
\begin{align} \label{eq:ZijDefn}
  \mu_{R}^{2}\frac{d}{d\mu_{R}^{2}}Z_{IJ} \equiv \gamma_{IK} Z_{KJ}\,
  \qquad \text{with} \qquad I,J,K={G,Q}
\end{align}
where $\gamma_{IK}$'s are the corresponding anomalous dimensions and
the summation over repeated index is understood.
The general solution to the RGE up to $a_{s}^{3}$ is obtained as
\begin{align}
  \label{eq:ZCoupSoln}
  Z_{IJ} &= \delta_{IJ} 
           + {a}_{s} \Bigg[ \frac{2}{\epsilon}
           \gamma_{IJ}^{(1)} \Bigg] 
           + {a}_{s}^{2} \Bigg[
           \frac{1}{\epsilon^{2}} \Bigg\{  2
           \beta_{0} \gamma_{IJ}^{(1)} + 2 \gamma_{IK}^{(1)} \gamma_{KJ}^{(1)}  \Bigg\} + \frac{1}{\epsilon} \Bigg\{ \gamma_{IJ}^{(2)}\Bigg\}
           \Bigg] 
           + {a}_{s}^{3} \Bigg[ \frac{1}{\epsilon^{3}} \Bigg\{
           \frac{8}{3} \beta_{0}^{2} \gamma_{IJ}^{(1)} 
           \nonumber\\
         &+ 4 \beta_{0} \gamma_{IK}^{(1)}
           \gamma_{KJ}^{(1)} + \frac{4}{3} \gamma_{IK}^{(1)} \gamma_{KL}^{(1)}
           \gamma_{LJ}^{(1)} \Bigg\} + \frac{1}{\epsilon^{2}} \Bigg\{ \frac{4}{3} \beta_{1} \gamma_{IJ}^{(1)} +
           \frac{4}{3} \beta_{0} \gamma_{IJ}^{(2)} 
           + \frac{2}{3}
           \gamma_{IK}^{(1)} \gamma_{KJ}^{(2)} 
          + \frac{4}{3} \gamma_{IK}^{(2)} \gamma_{KJ}^{(1)} \Bigg\} 
          \nonumber\\
         &+ \frac{1}{\epsilon}
           \Bigg\{ \frac{2}{3} \gamma_{IJ}^{(3)} \Bigg\} \Bigg]
\end{align}
where, $\gamma_{IJ}$ is expanded in powers of $a_{s}$ as
\begin{align}
  \label{eq:gammaijExp}
  \gamma_{IJ} = \sum_{n=1}^{\infty} a_{s}^{n} \gamma_{IJ}^{(n)}\,.
\end{align}
The second term of the Lagrangian can be written
in terms of renormalised quantities:
\begin{eqnarray}
-{1 \over 2} \int d^4x ~ h_{\mu\nu} \left(\kappa_G~ {\cal O}^{G,\mu\nu} + \kappa_Q~{\cal O}^{Q,\mu\nu} \right)
\end{eqnarray}
where the $\kappa_I$ are related to the bare ones by
\begin{eqnarray}
\hat \kappa_I = Z_{IJ} \kappa_J
\end{eqnarray}

\subsection{Infrared structure}  \label{sec:ff}
{
In the colour space, the matrix elements of 
unrenormalized composite operators $\hat {\cal O}^I,~ I=G,Q$ between a pair of on-shell partonic states $i=q,g$ and 
the vacuum state are expanded in powers of bare coupling constant $\hat a_s$ as 
\begin{equation}
 | \mathcal{M}^{I}_i \rangle = \sum_{n=0}^\infty \hat{a}_s^n \left({Q^2 \over \mu^2}\right)^{n \epsilon/2}
S^n_\epsilon| \uM^{I,(n)}_i \rangle
\end{equation}
where $i=q,\overline q, g$.
In terms of these, we can define the on-shell form factor of $\hat {\cal O}^I,~I=G,Q$ by taking the 
the overlap of $| \mathcal{M}^{I}_i \rangle$ with its leading order amplitude normalised
with respect to the leading order contribution.  Given these two operators, one finds
four independent form factors:  
\begin{equation}
 \hat{\cal F}^{I,g,(n)} = \frac{\langle \uM^{G,(0)}_g | \uM^{I,(n)}_g \rangle}{\langle \uM^{G,(0)}_g | \uM^{G,(0)}_g \rangle} \,,
 \hspace{1cm}
 \hat{\cal F}^{I,q,(n)} = \frac{\langle \uM^{Q,(0)}_q | \uM^{I,(n)}_q \rangle}{\langle \uM^{Q,(0)}_q | \uM^{Q,(0)}_q \rangle} \quad \quad \quad I=G,Q\,.
\end{equation}
Note that, the non-diagonal amplitudes \textit{i.e.} $| \uM^{Q,(n)}_g \rangle$ and $| \uM^{G,(n)}_q \rangle$, start at one-loop level
and hence, the corresponding form factors start at ${\cal O} (\hat{a}_s)$.

The form factors are often ill-defined in 4-dimensions even after UV renormalisation due the presence of infrared
divergences when massless modes are present.  
The massless gluons and light quarks and anti-quarks bring in these IR divergences beyond the leading
order in perturbation theory.  As we mentioned earlier, we regulate 
both UV and IR divergences using dimensional regularisation
without disturbing the gauge symmetry of the theory.     
The UV divergences are renormalised away by coupling constant as well as overall operator renormalizations.  The resulting UV finite form factors will contain
IR divergences which appear in terms of poles in $\epsilon$.  Thanks to factorisation properties and 
universality of these IR divergences, these on-shell form factors satisfy Sudakov differential equation,
famously known as K-G equation\footnote{The name is due to the presence of two functions in Sudakov differential equation which are popularly denoted by letters $K$ and $G$.}.  A generalisation to multiparton 
amplitudes up to two loop level in QCD      
was proposed by Catani~\cite{Catani:1998bh}  
using the universal IR di-pole subtraction operators,  see also \cite{Sterman:2002qn}. 
The generalisation of IR subtraction operators of Catani beyond two loops were proposed 
by Becher and Neubert~\cite{Becher:2009cu} and by Gardi and Magnea~\cite{Gardi:2009qi}. 
Following closely the notation used in~\cite{Ravindran:2005vv},
we find that the UV finite form factors ${{\cal F}}^{I,i} (\hat{a}_{s}, Q^{2}, \mu^{2}, \epsilon)$, 
after performing strong coupling constant and operator renormalizations,
satisfy the integro-differential K-G equation \cite{Sudakov:1954sw, Mueller:1979ih, Collins:1980ih, Sen:1981sd}
given by
\begin{equation} \label{eq:KG}
  Q^2 \frac{d}{dQ^2} \ln {\cal F}^{I,i} (\hat{a}_s, Q^2, \mu^2, \epsilon)
  = \frac{1}{2} \left[ K^{i} \left(\hat{a}_s, \frac{\mu_R^2}{\mu^2}, \epsilon\right) + G^{I,i} \left(\hat{a}_s, \frac{Q^2}{\mu_R^2}, \frac{\mu_R^2}{\mu^2}, \epsilon \right) \right]
\end{equation}
where the $Q^{2}=-q^2=-(p_1+p_2)^2$ with $p_i$ being the momenta of external on-shell states.   
The $Q^2$ independent function $K^{i}$ contains all the poles in the dimensional regulator $\epsilon$ and 
the terms, finite in $\epsilon \rightarrow 0$, are encapsulated in $G^{I,i}$. 

The solutions present a universal structure of the singularities, except the single pole in $\epsilon$.
Single poles are controlled by the finite functions $G^{I,i}$.  We find
\begin{eqnarray}
G^{I,i}\left(\hat a_s, {Q^2 \over \mu_R^2},{\mu_R^2 \over \mu^2},\epsilon\right)
        &=& G^{I,i}\left(a_s(\mu_R^2),{Q^2 \over \mu_R^2},\epsilon\right)  
\nonumber\\
        &=& G^{I,i} \left(a_s(Q^2), 1, \epsilon\right)+\int_{Q^2 \over \mu_R^2}^1 {d \lambda^2 \over \lambda^2}
            A^i(\lambda^2 \mu_R^2)
\end{eqnarray}
where $A^i$ are cusp anomalous dimension that do not depend on the type of operator $I$.

In \cite{Ravindran:2004mb, Moch:2005tm}, it was first observed that the 
coefficient $G^{I,i}$ of the single pole in $\epsilon$ 
manifests a universal structure, in terms of the anomalous dimensions. 
In  \cite{Ravindran:2004mb}, the factorization of the single
pole in quark and gluon form factors in terms of soft and collinear
anomalous dimensions was first revealed up
to two loop level whose validity at three loop was later established
in the article \cite{Moch:2005tm}.  That is, expanding $G^{I,i}$ as
\begin{eqnarray}
G^{I,i}\left(a_s(Q^2),1,\epsilon\right) = \sum_{n=1}^\infty a_s^n(Q^2) G^{I,i}_n(\epsilon)
\end{eqnarray}
one finds
\begin{align} \label{eq:GIi}
 G^{I,i}_n (\ep) = 2 B^i_n  + f^i_n + C^{I,i}_n + \sum_{k=1}^{\infty} \epsilon^k g_n^{I,i,k} \, ,
\end{align}
where, the constants $C_{n}^{I,i}$ up to three-loop are \cite{Ravindran:2006cg}
\begin{align} \label{eq:Cg}
C_{1}^{I,i} &= 0\, ,
\nonumber\\
C_{2}^{I,i} &= - 2 \beta_{0} g_{1}^{I,i,1}\, ,
\nonumber\\
C_{3}^{I,i} &= - 2 \beta_{1} g_{1}^{I,i,1} - 2 \beta_{0} \left(g_{2}^{I,i,1} + 2 \beta_{0} g_{1}^{I,i,2}\right)\, .
\end{align}
In the above expressions, $X^{I,i}_{n}$ with $X=A,B,f$
are defined through
\begin{align}  \label{eq:ABfgmExp}
  X^{I,i} &\equiv \sum_{n=1}^{\infty} a_{s}^{n}
                        X^{I,i}_{n}\,.
\end{align}
The constant $G^{I,i}_n (\ep)$ in Eq.~\ref{eq:GIi} 
depends not only on the universal collinear ($B^i_n$) and
soft ($f^i_n$) anomalous dimensions, but also the operator as well as process 
dependent constants $g_n^{I,i,k}$.   
In other words, except $g_n^{I,i,k}$, the solution to the K-G
equation contains only universal quantities such as $A^i,B^i$ and $f^i$,
in addition to $\beta_i$.  Since, $A^i$ \cite{Moch:2004pa, Vogt:2004mw,
Catani:1989ne, Catani:1990rp, Vogt:2000ci, Ahmed:2014cha} ,$B^i$
\cite{Vogt:2004mw} and $f^i$ \cite{Ravindran:2004mb, Moch:2005tm} are
known up to three loop level, we can use the solution to K-G equation 
to determine the renormalisation constants $Z_{IJ}$.  Hence our next
task is to compute the on-shell form factors order by order in
perturbation theory and compare them against the predictions of K-G
equation  to determine the unknown renormalisation constants $\gamma_{IJ}$
in $Z_{IJ}$.  Using these renormalisation constants, we obtain UV finite
on-shell form factors of ${\cal O}^I$ up to three loop level.  
\section{Computation and Results}
In this section, after a brief discussion on how we have performed the computation,
we present the unrenormalized form factors $\hat {\cal F}^{I,i,(n)}$ and 
the anomalous dimensions $\gamma_{IJ}$
up to three loop level.
We closely follow the steps used in the derivation of three loop 
unrenormalized form factors 
of scalar and vector form factors
\cite{Gehrmann:2010ue,Gehrmann:2014vha}, see also \cite{Ahmed:2015qia,Ahmed:2015qpa,Ahmed:2016vgl}.  The relevant Feynman diagrams 
are generated using
QGRAF~\cite{Nogueira:1991ex}. The numbers of diagrams contributing to
three loop amplitudes are 1586 for
$|{\hat{\cal M}}^{G,(3)}_{g}\rangle$, 447 for
$|{\hat{\cal M}}^{Q,(3)}_{g}\rangle$, 400 for
$|{\hat{\cal M}}^{G,(3)}_{q}\rangle$ and 244 for
$|{\hat{\cal M}}^{Q,(3)}_{q}\rangle$ where all the external particles
are considered to be on-shell. The QGRAF output is suitably 
converted to a format that can be further used to perform the
substitution of Feynman rules, contraction of Lorentz and colour indices 
and simplification of Dirac and Gell-Mann matrices.  We have used a 
set of in-house routines
written in the symbolic manipulating program FORM 
\cite{Vermaseren:2000nd} to achieve them.  We have included 
ghost loops in the Feynman gauge.  For the external on-shell gluons,
we have kept only transversely polarization states of gluons in n-dimensions.
The resulting large number of integrals are further reduced to  
fewer scalar integrals, called master integrals (MIs), using 
integration-by-parts (IBP) \cite{Tkachov:1981wb,Chetyrkin:1981qh} and
Lorentz invariance (LI) \cite{Gehrmann:1999as} identities.  
While the LI identities are not linearly independent from the IBP
identities~\cite{Lee:2008tj}, they however help to
accelerate to solve the large number of equations resulting from IBP. 
Reduction to MIs is achieved using Laporta algorithm,~\cite{Laporta:2001dd}
implemented in various symbolic manipulation packages such as 
AIR~\cite{Anastasiou:2004vj}, FIRE~\cite{Smirnov:2008iw},
Reduze2~\cite{vonManteuffel:2012np, Studerus:2009ye} and
LiteRed~\cite{Lee:2013mka, Lee:2012cn}.  
We first use Reduze2~\cite{vonManteuffel:2012np, Studerus:2009ye} to shift loop momenta to get suitable 
integral classes and then make extensive use of LiteRed~\cite{Lee:2013mka, Lee:2012cn} to perform
the reductions of all the integrals to MIs.  
We find that at three loop level, there are 22 topologically different master integrals (MIs) 
involving genuine three-loop
integrals with vertex functions ($A_{t,i}$), three-loop propagator
integrals ($B_{t,i}$) and products of one- and
two-loop integrals ($C_{t,i}$).  Each integral has been computed 
analytically as a Laurent series in $\epsilon$ and they can be found
in~\cite{Gehrmann:2005pd,Gehrmann:2006wg,Heinrich:2007at,Heinrich:2009be,Lee:2010cga}.
The entire set can also be found in the appendix of~\cite{Gehrmann:2010ue}.
Substituting these MIs, we obtain the unrenormalized form factors which are listed below: 
\begin{align}
\hat {\cal F}^{G,g,(0)} &= 1 \,,
\\
\hat {\cal F}^{G,g,(1)} &=
%
         C_A \Bigg\{ \frac{1}{\epsilon^{2}}   \Bigg(
          - 8
          \Bigg)
       + \frac{1}{\epsilon}   \Bigg(
            \frac{22}{3}
          \Bigg)
       +   \Bigg(
          - \frac{203}{18}
          + \zeta_2
          \Bigg)
       + \epsilon   \Bigg(
          + \frac{2879}{216}
          - \frac{7}{3} \zeta_3
          - \frac{11}{12} \zeta_2
          \Bigg)
\nonumber\\
&
       + \epsilon^2   \Bigg(
          - \frac{37307}{2592}
          + \frac{77}{36} \zeta_3
          + \frac{203}{144} \zeta_2
          + \frac{47}{80} \zeta_2^2
          \Bigg)
       + \epsilon^3   \Bigg(
            \frac{465143}{31104}
          - \frac{31}{20} \zeta_5
          - \frac{1421}{432} \zeta_3
\nonumber\\
&
           - \frac{2879}{1728} \zeta_2
          + \frac{7}{24} \zeta_2 \zeta_3
          - \frac{517}{960} \zeta_2^2
          \Bigg)
       + \epsilon^4   \Bigg(
          - \frac{5695811}{373248}
          + \frac{341}{240} \zeta_5
          + \frac{20153}{5184} \zeta_3
          - \frac{49}{144} \zeta_3^2
\nonumber\\
&          
         + \frac{37307}{20736} \zeta_2
          - \frac{77}{288} \zeta_2 \zeta_3
          + \frac{9541}{11520} \zeta_2^2
          + \frac{949}{4480} \zeta_2^3
          \Bigg)
     \Bigg\} \,,
\label{eq:FGg1}
\\
\hat {\cal F}^{G,g,(2)} &=
         C_F n_f  \Bigg\{  \frac{1}{\epsilon^{2}}   \Bigg(
            \frac{32}{9}
          \Bigg)
       + \frac{1}{\epsilon}   \Bigg(
          - \frac{260}{27}
          \Bigg)
       +   \Bigg(
            \frac{3037}{162}
          - \frac{8}{3} \zeta_2
          \Bigg)
       + \epsilon   \Bigg(
          - \frac{61807}{1944}
          + \frac{62}{27} \zeta_3
\nonumber\\
&
       + \frac{65}{9} \zeta_2
          \Bigg)
          + \epsilon^2   \Bigg(
            \frac{1158007}{23328}
          - \frac{461}{81} \zeta_3
          - \frac{3185}{216} \zeta_2
          - \frac{31}{45} \zeta_2^2
          \Bigg)
       + \epsilon^3   \Bigg(
          - \frac{20551495}{279936}
          - \frac{28}{45} \zeta_5
\nonumber\\
&          
          + \frac{26131}{1944} \zeta_3
          + \frac{22759}{864} \zeta_2
          - \frac{17}{6} \zeta_2 \zeta_3
          + \frac{1721}{1080} \zeta_2^2
          \Bigg)
%
       \Bigg\} 
       + C_A n_f   \Bigg\{  \frac{1}{\epsilon^{3}}   \Bigg(
          - \frac{8}{3}
          \Bigg)
       + \frac{1}{\epsilon^{2}}   \Bigg(
            \frac{64}{9}
          \Bigg)
  \nonumber\\
&          
          + \frac{1}{\epsilon}   \Bigg(
          - \frac{499}{27}
          + 2 \zeta_2
          \Bigg)
       +   \Bigg(
            \frac{6863}{162}
          - \frac{38}{9} \zeta_3
          - \frac{16}{3} \zeta_2
          \Bigg)
       + \epsilon   \Bigg(
          - \frac{84433}{972}
          + \frac{277}{27} \zeta_3
          + \frac{481}{36} \zeta_2
\nonumber\\
&
        \frac{73}{60} \zeta_2^2
          \Bigg)
       + \epsilon^2   \Bigg(
            \frac{1913059}{11664}
          - \frac{151}{30} \zeta_5
          - \frac{2269}{81} \zeta_3
          - \frac{1009}{36} \zeta_2
          + \frac{5}{2} \zeta_2 \zeta_3
          - \frac{131}{45} \zeta_2^2
          \Bigg)
 \nonumber\\
&          
          + \epsilon^3   \Bigg(
          - \frac{40845067}{139968}
          + \frac{559}{45} \zeta_5
          + \frac{251461}{3888} \zeta_3
          - \frac{343}{108} \zeta_3^2
          + \frac{68603}{1296} \zeta_2
          - \frac{25}{4} \zeta_2 \zeta_3
          + \frac{6911}{864} \zeta_2^2
   %
%
\nonumber\\
&
      + \frac{781}{1680} \zeta_2^3
          \Bigg)\Bigg\}  
       + C_A^2  \Bigg\{  \frac{1}{\epsilon^{4}}   \Bigg(
            32
          \Bigg)
       + \frac{1}{\epsilon^{3}}   \Bigg(
          - 44
          \Bigg)
       + \frac{1}{\epsilon^{2}}   \Bigg(
            \frac{226}{3}
          - 4 \zeta_2
          \Bigg)
       + \frac{1}{\epsilon}   \Bigg(
          - 81
\nonumber\\
&
        + \frac{50}{3} \zeta_3
          + \frac{11}{3} \zeta_2
          \Bigg)
       +   \Bigg(
            \frac{5249}{108}
          - 11 \zeta_3
          - \frac{67}{18} \zeta_2
          - \frac{21}{5} \zeta_2^2
          \Bigg)
       + \epsilon   \Bigg(
            \frac{59009}{1296}
          - \frac{71}{10} \zeta_5
          + \frac{433}{18} \zeta_3
 \nonumber\\
&        
          - \frac{337}{108} \zeta_2
          - \frac{23}{6} \zeta_2 \zeta_3
     + \frac{99}{40} \zeta_2^2
          \Bigg)
       + \epsilon^2   \Bigg(
          - \frac{1233397}{5184}
          + \frac{759}{20} \zeta_5
          - \frac{8855}{216} \zeta_3
          + \frac{901}{36} \zeta_3^2
  \nonumber\\
&   
          + \frac{12551}{648} \zeta_2
          + \frac{77}{36} \zeta_2 \zeta_3
        - \frac{4843}{720} \zeta_2^2
          + \frac{2313}{280} \zeta_2^3
          \Bigg)
       + \epsilon^3   \Bigg(
            \frac{108841321}{186624}
          - \frac{3169}{28} \zeta_7
          - \frac{4691}{60} \zeta_5
\nonumber\\
&   
          + \frac{22231}{216} \zeta_3
          - \frac{2365}{72} \zeta_3^2
          - \frac{813499}{15552} \zeta_2
          + \frac{313}{40} \zeta_2 \zeta_5
          - \frac{1609}{216} \zeta_2 \zeta_3
          + \frac{21901}{1440} \zeta_2^2
          - \frac{1291}{80} \zeta_2^2 \zeta_3
     \nonumber\\
&   
          - \frac{65659}{3360} \zeta_2^3
          \Bigg)
%
        \Bigg\} \,,
%
\\
\hat {\cal F}^{G,g,(3)} &=
         C_F n_f^2  \Bigg\{  \frac{1}{\epsilon^{3}}   \Bigg(
            \frac{256}{81}
          \Bigg)
       + \frac{1}{\epsilon^{2}}   \Bigg(
          - \frac{128}{9}
          \Bigg)
       + \frac{1}{\epsilon}   \Bigg(
            \frac{30916}{729}
          - \frac{160}{27} \zeta_2
          \Bigg)
       +   \Bigg(
          - \frac{78268}{729}
          + \frac{208}{81} \zeta_3
\nonumber\\
&            
          + \frac{80}{3} \zeta_2
          \Bigg)
       \Bigg\}
       + C_F^2 n_f  \Bigg\{  \frac{1}{\epsilon^{3}}   \Bigg(
            \frac{512}{81}
          \Bigg)
       + \frac{1}{\epsilon^{2}}   \Bigg(
          - \frac{1600}{81}
          \Bigg)
       + \frac{1}{\epsilon}   \Bigg(
            \frac{20180}{729}
          + \frac{320}{9} \zeta_3
          - \frac{320}{27} \zeta_2
          \Bigg)
\nonumber\\
&  
       +   \Bigg(
            \frac{35957}{2430}
          - \frac{45056}{405} \zeta_3
          + \frac{1144}{27} \zeta_2
          - \frac{32}{3} \zeta_2^2
          \Bigg)
       \Bigg\}
       + C_A n_f^2  \Bigg\{  \frac{1}{\epsilon^{4}}   \Bigg(
          - \frac{128}{81}
          \Bigg)
       + \frac{1}{\epsilon^{3}}   \Bigg(
            \frac{1696}{243}
          \Bigg)
\nonumber\\
&  
       + \frac{1}{\epsilon^{2}}   \Bigg(
          - \frac{6328}{243}
          + \frac{80}{27} \zeta_2
          \Bigg)
       + \frac{1}{\epsilon}   \Bigg(
            \frac{189167}{2187}
          - \frac{464}{81} \zeta_3
          - \frac{1060}{81} \zeta_2
          \Bigg)
       +   \Bigg(
          - \frac{6734887}{26244}
\nonumber\\
&            
          + \frac{5500}{243} \zeta_3
          + \frac{3805}{81} \zeta_2
          + \frac{293}{135} \zeta_2^2
          \Bigg)
       \Bigg\}
       + C_A C_F n_f  \Bigg\{  \frac{1}{\epsilon^{4}}   \Bigg(
          - \frac{256}{9}
          \Bigg)
       + \frac{1}{\epsilon^{3}}   \Bigg(
            \frac{2032}{27}
          \Bigg)
\nonumber\\
&  
       + \frac{1}{\epsilon^{2}}   \Bigg(
          - \frac{10532}{81}
          - \frac{64}{9} \zeta_3
          + \frac{224}{9} \zeta_2
          \Bigg)
       + \frac{1}{\epsilon}   \Bigg(
            \frac{39715}{243}
          - \frac{944}{27} \zeta_3
          - \frac{1490}{27} \zeta_2
          + \frac{32}{15} \zeta_2^2
          \Bigg)
\nonumber\\
&  
       +   \Bigg(
          - \frac{1315651}{14580}
          - \frac{112}{9} \zeta_5
          + \frac{29818}{405} \zeta_3
          + \frac{11719}{162} \zeta_2
          + \frac{40}{3} \zeta_2 \zeta_3
          + \frac{50}{3} \zeta_2^2
          \Bigg)
       \Bigg\}
\nonumber\\
&  
       + C_A^2 n_f  \Bigg\{  \frac{1}{\epsilon^{5}}   \Bigg(
            \frac{64}{3}
          \Bigg)
       + \frac{1}{\epsilon^{4}}   \Bigg(
          - \frac{4784}{81}
          \Bigg)
       + \frac{1}{\epsilon^{3}}   \Bigg(
            \frac{35764}{243}
          - \frac{376}{27} \zeta_2
          \Bigg)
       + \frac{1}{\epsilon^{2}}   \Bigg(
          - \frac{7435}{27}
\nonumber\\
&            
          + \frac{1208}{27} \zeta_3
          + \frac{2458}{81} \zeta_2
          \Bigg)
       + \frac{1}{\epsilon}   \Bigg(
            \frac{2991329}{8748}
          - \frac{6634}{81} \zeta_3
          - \frac{27059}{486} \zeta_2
          - \frac{1493}{90} \zeta_2^2
          \Bigg)
\nonumber\\
&  
       +   \Bigg(
            \frac{4440127}{524880}
          - \frac{3002}{45} \zeta_5
          + \frac{219163}{810} \zeta_3
          + \frac{229919}{5832} \zeta_2
          - \frac{331}{9} \zeta_2 \zeta_3
          + \frac{11461}{360} \zeta_2^2
          \Bigg)
       \Bigg\}
\nonumber\\
&  
       + C_A^3  \Bigg\{  \frac{1}{\epsilon^{6}}   \Bigg(
          - \frac{256}{3}
          \Bigg)
       + \frac{1}{\epsilon^{5}}   \Bigg(
            \frac{352}{3}
          \Bigg)
       + \frac{1}{\epsilon^{4}}   \Bigg(
          - \frac{14744}{81}
          \Bigg)
       + \frac{1}{\epsilon^{3}}   \Bigg(
            \frac{13126}{243}
          - \frac{176}{3} \zeta_3
\nonumber\\
&   
          + \frac{484}{27} \zeta_2
          \Bigg)
       + \frac{1}{\epsilon^{2}}   \Bigg(
            \frac{149939}{486}
          - \frac{440}{27} \zeta_3
          - \frac{4321}{81} \zeta_2
          + \frac{494}{45} \zeta_2^2
          \Bigg)
       + \frac{1}{\epsilon}   \Bigg(
          - \frac{14639165}{17496}
\nonumber\\
&            
          + \frac{1756}{15} \zeta_5
          - \frac{634}{9} \zeta_3
          + \frac{112633}{972} \zeta_2
          + \frac{170}{9} \zeta_2 \zeta_3
          + \frac{4213}{180} \zeta_2^2
          \Bigg)
       +   \Bigg(
            \frac{1056263429}{1049760}
          + \frac{5014}{45} \zeta_5
          \nonumber\\
&            
          + \frac{539}{2430} \zeta_3
          - \frac{1766}{9} \zeta_3^2
          - \frac{1988293}{11664} \zeta_2
          - \frac{92}{9} \zeta_2 \zeta_3
          - \frac{64997}{2160} \zeta_2^2
          - \frac{22523}{270} \zeta_2^3
          \Bigg)
       \Bigg\} \,,
\\
\hat {\cal F}^{Q,g,(0)} &= 0 \,,
%
\\
\hat {\cal F}^{Q,g,(1)} &=
         n_f  \Bigg\{  \frac{1}{\epsilon}   \Bigg(
          - \frac{4}{3}
          \Bigg)
       +   \Bigg(
            \frac{35}{18}
          \Bigg)
       + \epsilon   \Bigg(
          - \frac{497}{216}
          + \frac{1}{6} \zeta_2
          \Bigg)
       + \epsilon^2   \Bigg(
            \frac{6593}{2592}
          - \frac{7}{18} \zeta_3
          - \frac{35}{144} \zeta_2
          \Bigg)
\nonumber\\
& 
       + \epsilon^3   \Bigg(
          - \frac{84797}{31104}
          + \frac{245}{432} \zeta_3
          + \frac{497}{1728} \zeta_2
          + \frac{47}{480} \zeta_2^2
          \Bigg)
       + \epsilon^4   \Bigg(
            \frac{1072433}{373248}
          - \frac{31}{120} \zeta_5
          - \frac{3479}{5184} \zeta_3
\nonumber\\
&           
          - \frac{6593}{20736} \zeta_2
          + \frac{7}{144} \zeta_2 \zeta_3
          - \frac{329}{2304} \zeta_2^2
          \Bigg)
%
%
%
      \Bigg\} \,,
%
\\
\hat {\cal F}^{Q,g,(2)} &=
         C_F n_f  \Bigg\{  \frac{1}{\epsilon^{2}}   \Bigg(
          - \frac{32}{9}
          \Bigg)
       + \frac{1}{\epsilon}   \Bigg(
            \frac{206}{27}
          \Bigg)
       +   \Bigg(
          - \frac{695}{81}
          - 8 \zeta_3
          + \frac{8}{3} \zeta_2
          \Bigg)
       + \epsilon   \Bigg(
            \frac{149}{243}
          + \frac{469}{27} \zeta_3
\nonumber\\
&            
          - \frac{121}{18} \zeta_2
          + \frac{12}{5} \zeta_2^2
          \Bigg)
       + \epsilon^2   \Bigg(
            \frac{143693}{5832}
          - 14 \zeta_5
          - \frac{2554}{81} \zeta_3
          + \frac{1219}{108} \zeta_2
          + 2 \zeta_2 \zeta_3
          - \frac{95}{18} \zeta_2^2
          \Bigg)
\nonumber\\
&  
       + \epsilon^3   \Bigg(
          - \frac{1386569}{17496}
          + \frac{6037}{180} \zeta_5
          + \frac{104639}{1944} \zeta_3
          - \frac{23}{3} \zeta_3^2
          - \frac{6581}{432} \zeta_2
          - \frac{29}{12} \zeta_2 \zeta_3
          + \frac{20633}{2160} \zeta_2^2
\nonumber\\
&            
          + \frac{99}{35} \zeta_2^3
          \Bigg)
%
       \Bigg\}
       + C_A n_f  \Bigg\{  \frac{1}{\epsilon^{3}}   \Bigg(
            \frac{32}{3}
          \Bigg)
       + \frac{1}{\epsilon^{2}}   \Bigg(
          - \frac{184}{9}
          \Bigg)
       + \frac{1}{\epsilon}   \Bigg(
            \frac{868}{27}
          - \frac{8}{3} \zeta_2
          \Bigg)
       +   \Bigg(
          - \frac{15541}{324}
\nonumber\\
&            
          + \frac{128}{9} \zeta_3
          + \frac{53}{9} \zeta_2
          \Bigg)
       + \epsilon   \Bigg(
            \frac{273061}{3888}
          - \frac{823}{27} \zeta_3
          - \frac{649}{54} \zeta_2
          - \frac{61}{15} \zeta_2^2
          \Bigg)
       + \epsilon^2   \Bigg(
          - \frac{4764919}{46656}
\nonumber\\
&            
          + \frac{182}{15} \zeta_5
          + \frac{37373}{648} \zeta_3
          + \frac{14545}{648} \zeta_2
          - \frac{44}{9} \zeta_2 \zeta_3
          + \frac{541}{60} \zeta_2^2
          \Bigg)
       + \epsilon^3   \Bigg(
            \frac{83029021}{559872}
          - \frac{8507}{360} \zeta_5
\nonumber\\
&            
          - \frac{219191}{1944} \zeta_3
          + \frac{454}{27} \zeta_3^2
          - \frac{604667}{15552} \zeta_2
          + \frac{1307}{108} \zeta_2 \zeta_3
          - \frac{4783}{270} \zeta_2^2
          + \frac{109}{420} \zeta_2^3
          \Bigg)
%
       \Bigg\}  \,,
%
\\
\hat {\cal F}^{Q,g,(3)} &=
         C_F n_f^2  \Bigg\{  \frac{1}{\epsilon^{3}}   \Bigg(
          - \frac{256}{81}
          \Bigg)
       + \frac{1}{\epsilon^{2}}   \Bigg(
            \frac{112}{9}
          \Bigg)
       + \frac{1}{\epsilon^{1}}   \Bigg(
          - \frac{20440}{729}
          - \frac{32}{3} \zeta_3
          + \frac{160}{27} \zeta_2
          \Bigg)
       +   \Bigg(
            \frac{27661}{729}
\nonumber\\
&              
          + \frac{3500}{81} \zeta_3
          - 26 \zeta_2
          + \frac{16}{5} \zeta_2^2
          \Bigg)
       \Bigg\}
       + C_F^2 n_f  \Bigg\{  \frac{1}{\epsilon^{3}}   \Bigg(
          - \frac{512}{81}
          \Bigg)
       + \frac{1}{\epsilon^{2}}   \Bigg(
            \frac{1600}{81}
          \Bigg)
       + \frac{1}{\epsilon}   \Bigg(
          - \frac{19694}{729}
\nonumber\\
&            
          - \frac{320}{9} \zeta_3
          + \frac{320}{27} \zeta_2
          \Bigg)
       +   \Bigg(
          - \frac{34246}{1215}
          + 80 \zeta_5
          + \frac{25076}{405} \zeta_3
          - \frac{1144}{27} \zeta_2
          + \frac{32}{3} \zeta_2^2
          \Bigg)
       \Bigg\}
\nonumber\\
&  
       + C_A n_f^2  \Bigg\{  \frac{1}{\epsilon^{4}}   \Bigg(
            \frac{32}{9}
          \Bigg)
       + \frac{1}{\epsilon^{3}}   \Bigg(
          - \frac{1012}{81}
          \Bigg)
       + \frac{1}{\epsilon^{2}}   \Bigg(
            \frac{8029}{243}
          - \frac{28}{9} \zeta_2
          \Bigg)
       + \frac{1}{\epsilon}   \Bigg(
          - \frac{237197}{2916}
          + \frac{52}{3} \zeta_3
\nonumber\\
&            
          + \frac{235}{18} \zeta_2
          \Bigg)
       +   \Bigg(
            \frac{34159189}{174960}
          - \frac{59047}{810} \zeta_3
          - \frac{28457}{648} \zeta_2
          - \frac{983}{180} \zeta_2^2
          \Bigg)
       \Bigg\}
       + C_A C_F n_f  \Bigg\{  \frac{1}{\epsilon^{4}}   \Bigg(
            \frac{256}{9}
          \Bigg)
\nonumber\\
&  
       + \frac{1}{\epsilon^{3}}   \Bigg(
          - \frac{1648}{27}
          \Bigg)
       + \frac{1}{\epsilon^{2}}   \Bigg(
            \frac{5396}{81}
          + 64 \zeta_3
          - \frac{224}{9} \zeta_2
          \Bigg)
       + \frac{1}{\epsilon}   \Bigg(
          - \frac{4519}{243}
          - \frac{472}{9} \zeta_3
          + \frac{1418}{27} \zeta_2
\nonumber\\
&            
          - \frac{96}{5} \zeta_2^2
          \Bigg)
       +   \Bigg(
          - \frac{516221}{14580}
          + 152 \zeta_5
          - \frac{4508}{45} \zeta_3
          - \frac{9883}{162} \zeta_2
          - \frac{40}{3} \zeta_2 \zeta_3
          + \frac{146}{15} \zeta_2^2
          \Bigg)
       \Bigg\}
\nonumber\\
&  
       + C_A^2 n_f  \Bigg\{  \frac{1}{\epsilon^{5}}   \Bigg(
          - \frac{128}{3}
          \Bigg)
       + \frac{1}{\epsilon^{4}}   \Bigg(
            \frac{736}{9}
          \Bigg)
       + \frac{1}{\epsilon^{3}}   \Bigg(
          - \frac{9982}{81}
          + \frac{32}{3} \zeta_2
          \Bigg)
       + \frac{1}{\epsilon^{2}}   \Bigg(
            \frac{77047}{486}
          - \frac{296}{3} \zeta_3
\nonumber\\
&            
          - \frac{58}{3} \zeta_2
          \Bigg)
       + \frac{1}{\epsilon}   \Bigg(
          - \frac{96755}{648}
          + \frac{1385}{9} \zeta_3
          + \frac{115}{4} \zeta_2
          + \frac{147}{5} \zeta_2^2
          \Bigg)
       +   \Bigg(
          - \frac{1027661}{349920}
          - \frac{2842}{15} \zeta_5
\nonumber\\
&            
          - \frac{36668}{405} \zeta_3
          + \frac{1109}{432} \zeta_2
          + 37 \zeta_2 \zeta_3
          - \frac{4019}{90} \zeta_2^2
          \Bigg)
       \Bigg\} \,,
\\
\hat {\cal F}^{G,q,(0)} &= 0 \,,
%
\\
\hat {\cal F}^{G,q,(1)} &=
         C_F  \Bigg\{  \frac{1}{\epsilon}   \Bigg(
          - \frac{16}{3}
          \Bigg)
       +   \Bigg(
            \frac{34}{9}
          \Bigg)
       + \epsilon   \Bigg(
          - \frac{79}{27}
          + \frac{2}{3} \zeta_2
          \Bigg)
       + \epsilon^2   \Bigg(
            \frac{401}{162}
          - \frac{14}{9} \zeta_3
          - \frac{17}{36} \zeta_2
          \Bigg)
\nonumber\\
&  
       + \epsilon^3   \Bigg(
          - \frac{2179}{972}
          + \frac{119}{108} \zeta_3
          + \frac{79}{216} \zeta_2
          + \frac{47}{120} \zeta_2^2
          \Bigg)
       + \epsilon^4   \Bigg(
            \frac{12377}{5832}
          - \frac{31}{30} \zeta_5
          - \frac{553}{648} \zeta_3
          - \frac{401}{1296} \zeta_2
\nonumber\\
&            
          + \frac{7}{36} \zeta_2 \zeta_3
          - \frac{799}{2880} \zeta_2^2
          \Bigg)
%
%
%
       \Bigg\} \,,
%
\\
\hat {\cal F}^{G,q,(2)} &=
         C_F n_f  \Bigg\{  \frac{1}{\epsilon^{2}}   \Bigg(
          - \frac{64}{9}
          \Bigg)
       + \frac{1}{\epsilon}   \Bigg(
            \frac{376}{27}
          \Bigg)
       +   \Bigg(
          - \frac{1798}{81}
          + \frac{16}{9} \zeta_2
          \Bigg)
       + \epsilon   \Bigg(
            \frac{16259}{486}
          - \frac{256}{27} \zeta_3
\nonumber\\
&            
          - \frac{94}{27} \zeta_2
          \Bigg)
       + \epsilon^2   \Bigg(
          - \frac{289163}{5832}
          + \frac{1504}{81} \zeta_3
          + \frac{899}{162} \zeta_2
          + \frac{38}{15} \zeta_2^2
          \Bigg)
       + \epsilon^3   \Bigg(
            \frac{5125571}{69984}
          - \frac{544}{45} \zeta_5
\nonumber\\
&            
          - \frac{7192}{243} \zeta_3
          - \frac{16259}{1944} \zeta_2
          + \frac{64}{27} \zeta_2 \zeta_3
          - \frac{893}{180} \zeta_2^2
          \Bigg)
%
       \Bigg\}
       + C_F^2  \Bigg\{  \frac{1}{\epsilon^{3}}   \Bigg(
            \frac{128}{3}
          \Bigg)
       + \frac{1}{\epsilon^{2}}   \Bigg(
          - \frac{688}{9}
          \Bigg)
\nonumber\\
&  
       + \frac{1}{\epsilon}   \Bigg(
            \frac{3340}{27}
          - \frac{32}{3} \zeta_2
          \Bigg)
       +   \Bigg(
          - \frac{14257}{81}
          + \frac{224}{9} \zeta_3
          + \frac{236}{9} \zeta_2
          \Bigg)
       + \epsilon   \Bigg(
            \frac{229261}{972}
          - \frac{1012}{27} \zeta_3
\nonumber\\
&            
          - \frac{1211}{27} \zeta_2
          - \frac{28}{5} \zeta_2^2
          \Bigg)
       + \epsilon^2   \Bigg(
          - \frac{3597469}{11664}
          + \frac{248}{15} \zeta_5
          + \frac{5437}{81} \zeta_3
          + \frac{21233}{324} \zeta_2
          - \frac{56}{9} \zeta_2 \zeta_3
\nonumber\\
&            
          + \frac{743}{90} \zeta_2^2
          \Bigg)
       + \epsilon^3   \Bigg(
            \frac{56232181}{139968}
          - \frac{613}{45} \zeta_5
          - \frac{51995}{486} \zeta_3
          + \frac{196}{27} \zeta_3^2
          - \frac{350153}{3888} \zeta_2
          + \frac{461}{27} \zeta_2 \zeta_3
\nonumber\\
&            
          - \frac{3331}{216} \zeta_2^2
          - \frac{31}{21} \zeta_2^3
          \Bigg)
%
       \Bigg\}
       + C_A C_F  \Bigg\{  \frac{1}{\epsilon^{2}}   \Bigg(
            \frac{176}{9}
          \Bigg)
       + \frac{1}{\epsilon}   \Bigg(
          - \frac{1124}{27}
          \Bigg)
       +   \Bigg(
            \frac{5651}{81}
          - \frac{16}{9} \zeta_2
          \Bigg)
\nonumber\\
&  
       + \epsilon   \Bigg(
          - \frac{108275}{972}
          + \frac{356}{27} \zeta_3
          - \frac{86}{27} \zeta_2
          - \frac{16}{15} \zeta_2^2
          \Bigg)
       + \epsilon^2   \Bigg(
            \frac{2055287}{11664}
          - 24 \zeta_5
          - \frac{961}{162} \zeta_3
          + \frac{986}{81} \zeta_2
\nonumber\\
&            
          - \frac{16}{3} \zeta_2 \zeta_3
          - \frac{5}{6} \zeta_2^2
          \Bigg)
       + \epsilon^3   \Bigg(
          - \frac{38875571}{139968}
          + \frac{5377}{90} \zeta_5
          - \frac{110159}{1944} \zeta_3
          + \frac{88}{3} \zeta_3^2
          - \frac{47947}{1944} \zeta_2
\nonumber\\
&            
          + \frac{287}{27} \zeta_2 \zeta_3
          - \frac{5323}{1080} \zeta_2^2
          + \frac{484}{35} \zeta_2^3
          \Bigg)
%
       \Bigg\}  \,,
%
\\
\hat {\cal F}^{G,q,(3)} &=
         C_F n_f^2  \Bigg\{  \frac{1}{\epsilon^{3}}   \Bigg(
          - \frac{256}{27}
          \Bigg)
       + \frac{1}{\epsilon^{2}}   \Bigg(
            \frac{2464}{81}
          \Bigg)
       + \frac{1}{\epsilon}   \Bigg(
          - \frac{17216}{243}
          + \frac{32}{9} \zeta_2
          \Bigg)
       +   \Bigg(
            \frac{107816}{729}
          - \frac{800}{27} \zeta_3
\nonumber\\
&            
          - \frac{308}{27} \zeta_2
          \Bigg)
       \Bigg\}
       + C_F^2 n_f  \Bigg\{  \frac{1}{\epsilon^{4}}   \Bigg(
            \frac{640}{9}
          \Bigg)
       + \frac{1}{\epsilon^{3}}   \Bigg(
          - \frac{18544}{81}
          \Bigg)
       + \frac{1}{\epsilon^{2}}   \Bigg(
            \frac{130696}{243}
          - \frac{176}{9} \zeta_2
          \Bigg)
\nonumber\\
&  
       + \frac{1}{\epsilon}   \Bigg(
          - \frac{776510}{729}
          + \frac{752}{9} \zeta_3
          + \frac{706}{9} \zeta_2
          \Bigg)
       +   \Bigg(
            \frac{8387353}{4374}
          - \frac{18250}{81} \zeta_3
          - \frac{15461}{81} \zeta_2
          - \frac{289}{15} \zeta_2^2
          \Bigg)
       \Bigg\}
\nonumber\\
&  
       + C_F^3  \Bigg\{  \frac{1}{\epsilon^{5}}   \Bigg(
          - \frac{512}{3}
          \Bigg)
       + \frac{1}{\epsilon^{4}}   \Bigg(
            \frac{1472}{3}
          \Bigg)
       + \frac{1}{\epsilon^{3}}   \Bigg(
          - \frac{89312}{81}
          + 64 \zeta_2
          \Bigg)
       + \frac{1}{\epsilon^{2}}   \Bigg(
            \frac{55964}{27}
          - \frac{832}{3} \zeta_3
\nonumber\\
&           
          - \frac{1592}{9} \zeta_2
          \Bigg)
       + \frac{1}{\epsilon}   \Bigg(
          - \frac{2565953}{729}
          + \frac{2296}{3} \zeta_3
          + \frac{8644}{27} \zeta_2
          + \frac{1148}{15} \zeta_2^2
          \Bigg)
       +   \Bigg(
            \frac{16239107}{2916}
\nonumber\\
&             
          - \frac{656}{5} \zeta_5
          - \frac{162008}{81} \zeta_3
          - \frac{65755}{162} \zeta_2
          + \frac{440}{3} \zeta_2 \zeta_3
          - \frac{2451}{10} \zeta_2^2
          \Bigg)
       \Bigg\}
       + C_A C_F n_f  \Bigg\{  \frac{1}{\epsilon^{3}}   \Bigg(
            \frac{4928}{81}
          \Bigg)
\nonumber\\
& 
       + \frac{1}{\epsilon^{2}}   \Bigg(
          - \frac{51592}{243}
          \Bigg)
       + \frac{1}{\epsilon}   \Bigg(
            \frac{127238}{243}
          + \frac{256}{9} \zeta_3
          - \frac{280}{27} \zeta_2
          \Bigg)
       +   \Bigg(
          - \frac{2526404}{2187}
          + \frac{5960}{81} \zeta_3
\nonumber\\
&           
          - \frac{13}{27} \zeta_2
          - \frac{128}{9} \zeta_2^2
          \Bigg)
       \Bigg\}
       + C_A C_F^2  \Bigg\{  \frac{1}{\epsilon^{4}}   \Bigg(
          - \frac{704}{3}
          \Bigg)
       + \frac{1}{\epsilon^{3}}   \Bigg(
            \frac{21784}{27}
          - \frac{64}{3} \zeta_2
          \Bigg)
\nonumber\\
& 
       + \frac{1}{\epsilon^{2}}   \Bigg(
          - \frac{487996}{243}
          + \frac{416}{3} \zeta_3
          + \frac{352}{9} \zeta_2
          \Bigg)
       + \frac{1}{\epsilon}   \Bigg(
            \frac{3102511}{729}
          - \frac{6092}{9} \zeta_3
          - \frac{1321}{27} \zeta_2
          - \frac{536}{15} \zeta_2^2
          \Bigg)
\nonumber\\
& 
       +   \Bigg(
          - \frac{71606351}{8748}
          + \frac{1624}{3} \zeta_5
          + \frac{13865}{9} \zeta_3
          - \frac{7513}{162} \zeta_2
          - \frac{52}{3} \zeta_2 \zeta_3
          + \frac{14549}{90} \zeta_2^2
          \Bigg)
       \Bigg\}
\nonumber\\
& 
       + C_A^2 C_F  \Bigg\{  \frac{1}{\epsilon^{3}}   \Bigg(
          - \frac{7744}{81}
          \Bigg)
       + \frac{1}{\epsilon^{2}}   \Bigg(
            \frac{87352}{243}
          \Bigg)
       + \frac{1}{\epsilon}   \Bigg(
          - \frac{704276}{729}
          - \frac{128}{9} \zeta_3
          - \frac{88}{9} \zeta_2
          \Bigg)
\nonumber\\
& 
       +   \Bigg(
            \frac{5045099}{2187}
          - 240 \zeta_5
          + \frac{6098}{81} \zeta_3
          + 209 \zeta_2
          - \frac{104}{3} \zeta_2 \zeta_3
          + \frac{622}{15} \zeta_2^2
          \Bigg)
       \Bigg\} \,,
\\
\hat {\cal F}^{Q,q,(0)} &= 1 \,,
%
\\
\hat {\cal F}^{Q,q,(1)} &=
         C_F  \Bigg\{  \frac{1}{\epsilon^{2}}   \Bigg(
          - 8
          \Bigg)
       + \frac{1}{\epsilon^{1}}   \Bigg(
            \frac{34}{3}
          \Bigg)
       +   \Bigg(
          - \frac{124}{9}
          + \zeta_2
          \Bigg)
       + \epsilon   \Bigg(
            \frac{403}{27}
          - \frac{7}{3} \zeta_3
          - \frac{17}{12} \zeta_2
          \Bigg)
\nonumber\\
& 
       + \epsilon^2   \Bigg(
          - \frac{2507}{162}
          + \frac{119}{36} \zeta_3
          + \frac{31}{18} \zeta_2
          + \frac{47}{80} \zeta_2^2
          \Bigg)
       + \epsilon^3   \Bigg(
            \frac{15301}{972}
          - \frac{31}{20} \zeta_5
          - \frac{217}{54} \zeta_3
          - \frac{403}{216} \zeta_2
\nonumber\\
&           
          + \frac{7}{24} \zeta_2 \zeta_3
          - \frac{799}{960} \zeta_2^2
          \Bigg)
       + \epsilon^4   \Bigg(
          - \frac{92567}{5832}
          + \frac{527}{240} \zeta_5
          + \frac{2821}{648} \zeta_3
          - \frac{49}{144} \zeta_3^2
          + \frac{2507}{1296} \zeta_2
\nonumber\\
&           
          - \frac{119}{288} \zeta_2 \zeta_3
          + \frac{1457}{1440} \zeta_2^2
          + \frac{949}{4480} \zeta_2^3
          \Bigg)
      \Bigg\}  \,,
%
\\
\hat {\cal F}^{Q,q,(2)} &=
         C_F n_f  \Bigg\{  \frac{1}{\epsilon^{3}}   \Bigg(
          - \frac{8}{3}
          \Bigg)
       + \frac{1}{\epsilon^{2}}   \Bigg(
            \frac{40}{3}
          \Bigg)
       + \frac{1}{\epsilon}   \Bigg(
          - \frac{89}{3}
          - \frac{2}{3} \zeta_2
          \Bigg)
       +   \Bigg(
            \frac{1909}{36}
          - \frac{26}{9} \zeta_3
          + \frac{22}{9} \zeta_2
          \Bigg)
\nonumber\\
& 
       + \epsilon   \Bigg(
          - \frac{36925}{432}
          + \frac{86}{9} \zeta_3
          - \frac{613}{108} \zeta_2
          + \frac{41}{60} \zeta_2^2
          \Bigg)
       + \epsilon^2   \Bigg(
            \frac{677941}{5184}
          - \frac{121}{30} \zeta_5
          - \frac{2317}{108} \zeta_3
\nonumber\\
&           
          + \frac{15745}{1296} \zeta_2
          - \frac{13}{18} \zeta_2 \zeta_3
          - \frac{359}{180} \zeta_2^2
          \Bigg)
       + \epsilon^3   \Bigg(
          - \frac{12131053}{62208}
          + \frac{67}{6} \zeta_5
          + \frac{52237}{1296} \zeta_3
          - \frac{169}{108} \zeta_3^2
\nonumber\\
&           
          - \frac{364273}{15552} \zeta_2
          + \frac{209}{54} \zeta_2 \zeta_3
          + \frac{19369}{4320} \zeta_2^2
          + \frac{127}{112} \zeta_2^3
          \Bigg)
%
       \Bigg\}
       + C_F^2  \Bigg\{  \frac{1}{\epsilon^{4}}   \Bigg(
            32
          \Bigg)
       + \frac{1}{\epsilon^{3}}   \Bigg(
          - \frac{272}{3}
          \Bigg)
\nonumber\\
& 
       + \frac{1}{\epsilon^{2}}   \Bigg(
            \frac{1570}{9}
          - 8 \zeta_2
          \Bigg)
       + \frac{1}{\epsilon}   \Bigg(
          - \frac{15023}{54}
          + \frac{128}{3} \zeta_3
          + \frac{32}{3} \zeta_2
          \Bigg)
       +   \Bigg(
            \frac{257615}{648}
          - \frac{1034}{9} \zeta_3
\nonumber\\
&           
          - \frac{103}{18} \zeta_2
          - 13 \zeta_2^2
          \Bigg)
       + \epsilon   \Bigg(
          - \frac{4112375}{7776}
          + \frac{92}{5} \zeta_5
          + \frac{13967}{54} \zeta_3
          - \frac{3767}{216} \zeta_2
          - \frac{56}{3} \zeta_2 \zeta_3
\nonumber\\
&           
          + \frac{1033}{30} \zeta_2^2
          \Bigg)
       + \epsilon^2   \Bigg(
            \frac{62375663}{93312}
          - \frac{1429}{30} \zeta_5
          - \frac{356111}{648} \zeta_3
          + \frac{652}{9} \zeta_3^2
          + \frac{177023}{2592} \zeta_2
          + \frac{691}{18} \zeta_2 \zeta_3
\nonumber\\
&           
          - \frac{56369}{720} \zeta_2^2
          + \frac{223}{20} \zeta_2^3
          \Bigg)
       + \epsilon^3   \Bigg(
          - \frac{911224295}{1119744}
          - \frac{4471}{28} \zeta_7
          + \frac{9439}{72} \zeta_5
          + \frac{8942747}{7776} \zeta_3
\nonumber\\
&           
          - \frac{21385}{108} \zeta_3^2
          - \frac{5072471}{31104} \zeta_2
          - \frac{23}{5} \zeta_2 \zeta_5
          - \frac{16141}{216} \zeta_2 \zeta_3
          + \frac{488237}{2880} \zeta_2^2
          - \frac{686}{15} \zeta_2^2 \zeta_3
%
  %
\nonumber\\
& 
       - \frac{3001}{105} \zeta_2^3
          \Bigg)
       \Bigg\}
    + C_A C_F  \Bigg\{  \frac{1}{\epsilon^{3}}   \Bigg(
      \frac{44}{3}
          \Bigg)
       + \frac{1}{\epsilon^{2}}   \Bigg(
          - \frac{508}{9}
          + 4 \zeta_2
          \Bigg)
       + \frac{1}{\epsilon}   \Bigg(
            \frac{7169}{54}
          - 26 \zeta_3
          + \frac{11}{3} \zeta_2
          \Bigg)
\nonumber\\
& 
       +   \Bigg(
          - \frac{165413}{648}
          + \frac{755}{9} \zeta_3
          - \frac{235}{9} \zeta_2
          + \frac{44}{5} \zeta_2^2
          \Bigg)
       + \epsilon   \Bigg(
            \frac{3429125}{7776}
          - \frac{51}{2} \zeta_5
          - \frac{5629}{27} \zeta_3
\nonumber\\
&           
          + \frac{15449}{216} \zeta_2
          + \frac{89}{6} \zeta_2 \zeta_3
          - \frac{1057}{40} \zeta_2^2
          \Bigg)
       + \epsilon^2   \Bigg(
          - \frac{66913709}{93312}
          + \frac{5411}{60} \zeta_5
          + \frac{286661}{648} \zeta_3
\nonumber\\
&   
          - \frac{569}{12} \zeta_3^2
        - \frac{383285}{2592} \zeta_2
          - \frac{877}{36} \zeta_2 \zeta_3
          + \frac{2527}{40} \zeta_2^2
          - \frac{809}{280} \zeta_2^3
          \Bigg)
       + \epsilon^3   \Bigg(
            \frac{1260896789}{1119744}
          + \frac{93}{2} \zeta_7
\nonumber\\
&            
          - \frac{42157}{180} \zeta_5
           - \frac{6822089}{7776} \zeta_3
          + \frac{29399}{216} \zeta_3^2
          + \frac{8369333}{31104} \zeta_2
          + \frac{497}{40} \zeta_2 \zeta_5
          + \frac{3683}{108} \zeta_2 \zeta_3
 \nonumber\\
&            
          - \frac{1142729}{8640} \zeta_2^2
           + \frac{7103}{240} \zeta_2^2 \zeta_3
          - \frac{143}{160} \zeta_2^3
          \Bigg)
%
       \Bigg\} \,,
%
\\
\hat {\cal F}^{Q,q,(3)} &=
         C_F n_f^2  \Bigg\{  \frac{1}{\epsilon^{4}}   \Bigg(
          - \frac{128}{81}
          \Bigg)
       + \frac{1}{\epsilon^{3}}   \Bigg(
            \frac{3808}{243}
          \Bigg)
       + \frac{1}{\epsilon^{2}}   \Bigg(
          - \frac{4240}{81}
          - \frac{16}{9} \zeta_2
          \Bigg)
\nonumber\\
&
       + \frac{1}{\epsilon}   \Bigg(
            \frac{283256}{2187}
          - \frac{272}{81} \zeta_3
          + \frac{284}{27} \zeta_2
          \Bigg)
       +   \Bigg(
          - \frac{1827880}{6561}
          + \frac{4348}{243} \zeta_3
          - \frac{314}{9} \zeta_2
         - \frac{83}{135} \zeta_2^2
          \Bigg)
       \Bigg\}
 \nonumber\\
&
       + C_F^2 n_f  \Bigg\{  \frac{1}{\epsilon^{5}}   \Bigg(
            \frac{64}{3}
          \Bigg)
       + \frac{1}{\epsilon^{4}}   \Bigg(
          - \frac{1232}{9}
          \Bigg)
       + \frac{1}{\epsilon^{3}}   \Bigg(
            \frac{33784}{81}
          + \frac{8}{3} \zeta_2
          \Bigg)
       + \frac{1}{\epsilon^{2}}   \Bigg(
          - \frac{232876}{243}
\nonumber\\
&          
          + \frac{584}{9} \zeta_3
          - \frac{94}{3} \zeta_2
          \Bigg)
       + \frac{1}{\epsilon}   \Bigg(
            \frac{1359371}{729}
          - \frac{8234}{27} \zeta_3
          + \frac{3533}{27} \zeta_2
          - \frac{337}{18} \zeta_2^2
          \Bigg)
       +   \Bigg(
          - \frac{28437107}{8748}
\nonumber\\
&          
          + \frac{278}{45} \zeta_5
          + \frac{3287}{3} \zeta_3
          - \frac{849}{2} \zeta_2
          - \frac{343}{9} \zeta_2 \zeta_3
          + \frac{69809}{1080} \zeta_2^2
          \Bigg)
       \Bigg\}
       + C_F^3  \Bigg\{  \frac{1}{\epsilon^{6}}   \Bigg(
          - \frac{256}{3}
          \Bigg)
\nonumber\\
&
       + \frac{1}{\epsilon^{5}}   \Bigg(
            \frac{1088}{3}
          \Bigg)
       + \frac{1}{\epsilon^{4}}   \Bigg(
          - \frac{2864}{3}
          + 32 \zeta_2
          \Bigg)
       + \frac{1}{\epsilon^{3}}   \Bigg(
            \frac{161240}{81}
          - \frac{800}{3} \zeta_3
          - 40 \zeta_2
          \Bigg)
\nonumber\\
&
       + \frac{1}{\epsilon^{2}}   \Bigg(
          - \frac{97202}{27}
          + \frac{3256}{3} \zeta_3
          - \frac{730}{9} \zeta_2
          + \frac{426}{5} \zeta_2^2
          \Bigg)
       + \frac{1}{\epsilon}   \Bigg(
            \frac{8625031}{1458}
          - \frac{1288}{5} \zeta_5
\nonumber\\
&          
          - 3050 \zeta_3
          + \frac{15017}{27} \zeta_2
          + \frac{428}{3} \zeta_2 \zeta_3
          - \frac{633}{2} \zeta_2^2
          \Bigg)
       +   \Bigg(
          - \frac{53150197}{5832}
          + \frac{14042}{15} \zeta_5
          + \frac{590021}{81} \zeta_3
\nonumber\\
&          
          - \frac{1826}{3} \zeta_3^2
          - \frac{576475}{324} \zeta_2
          - 267 \zeta_2 \zeta_3
          + \frac{289927}{360} \zeta_2^2
          - \frac{9095}{252} \zeta_2^3
          \Bigg)
       \Bigg\}
       + C_A C_F n_f  \Bigg\{  \frac{1}{\epsilon^{4}}   \Bigg(
            \frac{1408}{81}
          \Bigg)
\nonumber\\
&
       + \frac{1}{\epsilon^{3}}   \Bigg(
          - \frac{32816}{243}
          + \frac{128}{27} \zeta_2
          \Bigg)
       + \frac{1}{\epsilon^{2}}   \Bigg(
            \frac{12868}{27}
          - \frac{1024}{27} \zeta_3
          + \frac{1264}{81} \zeta_2
          \Bigg)
       + \frac{1}{\epsilon}   \Bigg(
          - \frac{2758264}{2187}
\nonumber\\
&          
          + \frac{17480}{81} \zeta_3
          - \frac{38542}{243} \zeta_2
          + \frac{88}{5} \zeta_2^2
          \Bigg)
       +   \Bigg(
            \frac{18919184}{6561}
          - \frac{128}{3} \zeta_5
          - \frac{70690}{81} \zeta_3
          + \frac{916919}{1458} \zeta_2
\nonumber\\
&          
          + \frac{392}{9} \zeta_2 \zeta_3
          - \frac{1777}{27} \zeta_2^2
          \Bigg)
       \Bigg\}
       + C_A C_F^2  \Bigg\{  \frac{1}{\epsilon^{5}}   \Bigg(
          - \frac{352}{3}
          \Bigg)
       + \frac{1}{\epsilon^{4}}   \Bigg(
            \frac{5560}{9}
          - 32 \zeta_2
          \Bigg)
\nonumber\\
&
       + \frac{1}{\epsilon^{3}}   \Bigg(
          - \frac{51404}{27}
          + 208 \zeta_3
          + \frac{92}{3} \zeta_2
          \Bigg)
       + \frac{1}{\epsilon^{2}}   \Bigg(
            \frac{1110322}{243}
          - \frac{3704}{3} \zeta_3
          + \frac{2119}{9} \zeta_2
          - \frac{332}{5} \zeta_2^2
          \Bigg)
\nonumber\\
&
       + \frac{1}{\epsilon}   \Bigg(
          - \frac{13792217}{1458}
          + 284 \zeta_5
          + \frac{37901}{9} \zeta_3
          - \frac{68459}{54} \zeta_2
          - \frac{430}{3} \zeta_2 \zeta_3
          + \frac{72523}{180} \zeta_2^2
          \Bigg)
\nonumber\\
&
       +   \Bigg(
            \frac{311359573}{17496}
          - \frac{42634}{45} \zeta_5
          - \frac{23739}{2} \zeta_3
          + \frac{1616}{3} \zeta_3^2
          + \frac{1339027}{324} \zeta_2
          + \frac{2026}{9} \zeta_2 \zeta_3
\nonumber\\
&          
          - \frac{2603779}{2160} \zeta_2^2
          - \frac{18619}{1260} \zeta_2^3
          \Bigg)
       \Bigg\}
       + C_A^2 C_F  \Bigg\{  \frac{1}{\epsilon^{4}}   \Bigg(
          - \frac{3872}{81}
          \Bigg)
       + \frac{1}{\epsilon^{3}}   \Bigg(
            \frac{75400}{243}
          - \frac{704}{27} \zeta_2
          \Bigg)
\nonumber\\
&
       + \frac{1}{\epsilon^{2}}   \Bigg(
          - \frac{10172}{9}
          + \frac{6688}{27} \zeta_3
          - \frac{2212}{81} \zeta_2
          - \frac{352}{45} \zeta_2^2
          \Bigg)
       + \frac{1}{\epsilon}   \Bigg(
            \frac{6969164}{2187}
          + \frac{272}{3} \zeta_5
          - \frac{36500}{27} \zeta_3
\nonumber\\
&          
          + \frac{123145}{243} \zeta_2
          + \frac{176}{9} \zeta_2 \zeta_3
          - \frac{1604}{15} \zeta_2^2
          \Bigg)
       +   \Bigg(
          - \frac{102217595}{13122}
          - \frac{428}{9} \zeta_5
          + \frac{2427625}{486} \zeta_3
\nonumber\\
&          
          - \frac{1136}{9} \zeta_3^2
          - \frac{1632292}{729} \zeta_2
          - \frac{614}{9} \zeta_2 \zeta_3
          + \frac{247963}{540} \zeta_2^2
          - \frac{6152}{189} \zeta_2^3
          \Bigg)
       \Bigg\}  \,.
\label{eq:FQq3}
\end{align}
where $C_A = N$ and $C_F = (N^2-1)/2N$ are the quadratic Casimir of
the $SU(N)$ group.  $T_F = 1/2$ and $n_f$ is the number of light active
quark flavours.  $\zeta_i$ is the Riemann Zeta function.

Having computed the unrenormalized form factors, our next task is to determine
the operator renormalisation constants $Z_{IJ}$.  As we explained in the previous section,
we can determine them by exploiting the universal IR structure of the form factors.  We determine these constants 
by comparing order by order the results of renormalised form factors expressed in terms of unknown $\gamma_{IJ}$  
against the predictions of the K-G equation expressed in terms of     
$A^i$, $B^i$ and $f^i$ anomalous dimensions that are known to three loop level.
The $\gamma_{IJ}$ thus extracted are listed below:
\begin{align}
 \gamma_{GG}^{(1)} &= - \frac{2}{3} n_f\,
 \\
 \gamma_{GG}^{(2)} &= - \frac{35}{27} C_A n_f - \frac{74}{27} C_F n_f
 \\
 \gamma_{GG}^{(3)} &=  C_A^2 n_f \Bigg( - \frac{3589}{162} + 24 \zeta_{3} \Bigg) + C_A C_F n_f \Bigg( \frac{139}{9} - \frac{104}{3} \zeta_{3} \Bigg) 
                + C_F^2 n_f \Bigg( - \frac{2155}{243} + \frac{32}{3} \zeta_{3} \Bigg) 
 \nonumber\\
 &
                + C_A n_f^2 \Bigg( \frac{1058}{243} \Bigg) - C_F n_f^2 \Bigg( \frac{173}{243} \Bigg)
 \\
 \gamma_{GQ}^{(1)} &=C_F \Bigg( \frac{8}{3} \Bigg)
 \\
 \gamma_{GQ}^{(2)} &= C_A C_F \Bigg( \frac{376}{27}  \Bigg)- C_F^2 \Bigg( \frac{112}{27} \Bigg) - C_F n_f \Bigg( \frac{104}{27} \Bigg)
 \\ 
 \gamma_{GQ}^{(3)} &= C_A^2 C_F \Bigg( \frac{20920}{243} + \frac{64}{3} \zeta_3 \Bigg) + C_A C_F^2 \Bigg( - \frac{8528}{243} - 64 \zeta_{3}\Bigg) 
                + C_F^3 \Bigg( - \frac{560}{243} + \frac{128}{3} \zeta_{3} \Bigg) 
 \nonumber\\
 &
                + C_A C_F n_f \Bigg( - \frac{22}{9} - \frac{128}{3} \zeta_{3} \Bigg) 
                + C_F^2 n_f \Bigg( -\frac{7094}{243} + \frac{128}{3} \zeta_{3} \Bigg) - C_F n_f^2 \Bigg( \frac{284}{81} \Bigg)
\end{align}
The remaining entries are
$ \gamma_{QG}^{(n)} = - \gamma_{GG}^{(n)}$ and  $\gamma_{QQ}^{(n)} = - \gamma_{GQ}^{(n)}$  where $n=1,2,3$.  This is indeed
a consequence of the fact that the sum of these operators is conserved.  This provides a crucial check on the correctness of our
computation.  Note that, all the $\gamma_{GG}^{(n)}$ are proportional to $n_{f}$ which is consistent
with the expectation that the conservation property of the operator $\hat {\cal O}^{G,\mu \nu}$ breaks down
beyond tree level due to the presence of quark loops.

The renormalised form factors can be obtained using Eq.(\ref{eq:asAasc},\ref{eq:Zas},\ref{eq:Zmat},\ref{eq:ZCoupSoln}).  Setting $\mu_R^2=Q^2$,  expanding in terms of $a_s(Q^2)$ as   
\begin{eqnarray}
F^{I,i} (Q^2) = \sum_{n=0}^\infty a_s^n(Q^2) {\cal F}^{I,i,(n)},
\quad \quad I=G,Q \quad \quad  i=g,q.
\end{eqnarray}
where $F^{I,i,(n)}$ up to three loop level are given by 

\begin{align}
\mathcal{F}^{G,g,(1)} =&  \frac{2}{\epsilon}\gamma^{(1)}_{GG} + \hat{\mathcal{F}}^{G,g,(1)} 
\nonumber\\
\mathcal{F}^{G,g,(2)} =& \frac{2}{\epsilon^2}\Big\{\beta_0 \gamma^{(1)}_{GG} + (\gamma^{(1)}_{GG})^2 + \gamma^{(1)}_{GQ}\gamma^{(1)}_{QG}\Big\} + \frac{1}{\epsilon}
\Big\{2\hat{\mathcal{F}}^{G,g,(1)}(\beta_0 + \gamma^{(1)}_{GG})
\nonumber\\
&2\hat{\mathcal{F}}^{Q,g,(1)}\gamma^{(1)}_{GQ} + \gamma^{(2)}_{GG}\Big\} + \hat{\mathcal{F}}^{G,g,(2)}
\nonumber\\
\mathcal{F}^{G,g,(3)} =&\frac{1}{\epsilon^3}\Big\{ \frac{8}{3}\beta_0^2\gamma^{(1)}_{GG} + 4\beta_0(\gamma^{(1)}_{GG})^2 + \frac{4}{3} (\gamma^{(1)}_{GG})^3 + 4\beta_0\gamma^{(1)}_{GQ}\gamma^{(1)}_{QG} + \frac{8}{3}\gamma^{(1)}_{GG}\gamma^{(1)}_{GQ}\gamma^{(1)}_{QG}
\nonumber\\
 &+\frac{4}{3}\gamma^{(1)}_{GQ}\gamma^{(1)}_{QQ}\gamma^{(1)}_{QG} \Big\} + \frac{1}{\epsilon^2}\Big\{4\beta_0^2\hat{\mathcal{F}}^{G,g,(1)} + \frac{4}{3}\beta_1\gamma^{(1)}_{GG} + 6\beta_0\hat{\mathcal{F}}^{G,g,(1)}\gamma^{(1)}_{GG}   
 \nonumber\\
 & +2\hat{\mathcal{F}}^{G,g,(1)} \big(\gamma^{(1)}_{GG}\big)^2 + 6\beta_0\hat{\mathcal{F}}^{Q,g,(1)}\gamma^{(1)}_{GQ} + 2\hat{\mathcal{F}}^{Q,g,(1)}\gamma^{(1)}_{GG}\gamma^{(1)}_{GQ}  + \hat{\mathcal{F}}^{G,g,(1)}\gamma^{(1)}_{GQ}\gamma^{(1)}_{QG} 
 \nonumber\\
 &+ \hat{\mathcal{F}}^{Q,g,(1)}\gamma^{(1)}_{GQ}\gamma^{(1)}_{QQ} + \frac{4}{3} \beta_0\gamma^{(2)}_{GG} + 2\gamma^{(1)}_{GG}\gamma^{(2)}_{GG}+\frac{4}{3}\gamma^{(2)}_{GQ} \gamma^{(1)}_{QG}+\frac{2}{3}\gamma^{(1)}_{GQ}\gamma^{(2)}_{QG}\Big\} 
 \nonumber\\
 &+ \frac{1}{\epsilon}\Big\{ \beta_1\hat{\mathcal{F}}^{G,g,(1)} + 4\beta_0\hat{\mathcal{F}}^{G,g,(2)} + 2\hat{\mathcal{F}}^{G,g,(2)}\gamma^{(1)}_{GG} + 2\hat{\mathcal{F}}^{Q,g,(2)}\gamma^{(1)}_{GQ} + \hat{\mathcal{F}}^{G,g,(1)}\gamma^{(2)}_{GG}  
 \nonumber\\
 &+\hat{\mathcal{F}}^{Q,g,(1)}\gamma^{(2)}_{GQ} + \frac{2}{3}\gamma^{(3)}_{GG} \Big\} + \hat{\mathcal{F}}^{G,g,(3)}
\nonumber\\
\mathcal{F}^{G,q,(1)} =& \frac{2}{\epsilon} \gamma^{(1)}_{GQ} + \hat{\mathcal{F}}^{G,q,(1)} 
\nonumber\\
 \mathcal{F}^{G,q,(2)} = &\frac{2}{\epsilon^2}\Big\{ \beta_0\gamma^{(1)}_{GQ} +\gamma^{(1)}_{GG}\gamma^{(1)}_{GQ} + \gamma^{(1)}_{GQ}\gamma^{(1)}_{QQ}  \Big\} + \frac{1}{\epsilon}\Big\{ 2\beta_0\hat{\mathcal{F}}^{G,q,(1)} + 2\hat{\mathcal{F}}^{G,q,(1)}\gamma^{(1)}_{GG} 
 \nonumber\\
 &+ 2\hat{\mathcal{F}}^{Q,q,(1)} \gamma^{(1)}_{GQ}   + \gamma^{(2)}_{GQ}  \Big\} + \hat{\mathcal{F}}^{G,q,(2)}
 \nonumber\\
 \mathcal{F}^{G,q,(3)} = &\frac{1}{\epsilon^3}\Big\{ \frac{8}{3}\beta_0^2\gamma^{(1)}_{GQ} + 4\beta_0\gamma^{(1)}_{GG} \gamma^{(1)}_{GQ}+ \frac{4}{3} (\gamma^{(1)}_{GG})^2\gamma^{(1)}_{GQ} + \frac{4}{3}(\gamma^{(1)}_{GQ})^2\gamma^{(1)}_{QG}  + 4\beta_0\gamma^{(1)}_{GQ}\gamma^{(1)}_{QQ}
 \nonumber\\
 &+\frac{4}{3}\gamma^{(1)}_{GG}\gamma^{(1)}_{GQ}\gamma^{(1)}_{QQ} + \frac{4}{3}\gamma^{(1)}_{GQ}(\gamma^{(1)}_{QQ})^2 \Big\} + \frac{1}{\epsilon^2}\Big\{ 4\beta_0^2\hat{\mathcal{F}}^{G,q,(1)}+ 6\beta_0\hat{\mathcal{F}}^{G,q,(1)}\gamma^{(1)}_{GG} 
 \nonumber\\
 &+ 2\hat{\mathcal{F}}^{G,q,(1)}(\gamma^{(1)}_{GG})^2  + \frac{4}{3}\beta_1\gamma^{(1)}_{GQ} + 6\beta_0\hat{\mathcal{F}}^{Q,q,(1)}\gamma^{(1)}_{GQ} + 2\hat{\mathcal{F}}^{Q,q,(1)}\gamma^{(1)}_{GG}\gamma^{(1)}_{GQ} 
 \nonumber\\
 &+ 2\hat{\mathcal{F}}^{G,q,(1)}\gamma^{(1)}_{GQ}\gamma^{(1)}_{QG} + 2\hat{\mathcal{F}}^{Q,q,(1)}\gamma^{(1)}_{GQ}\gamma^{(1)}_{QQ} + \frac{4}{3}\gamma^{(1)}_{GQ}\gamma^{(2)}_{GG} + \frac{4}{3}\beta_0\gamma^{(2)}_{GQ} + \frac{2}{3}\gamma^{(1)}_{GG}\gamma^{(2)}_{GQ}
 \nonumber\\
 & + \frac{4}{3}\gamma^{(1)}_{QQ}\gamma^{(2)}_{GQ} + \frac{2}{3}\gamma^{(1)}_{GQ}\gamma^{(2)}_{QQ} \Big\}+\frac{1}{\epsilon}\Big\{  \beta_1\hat{\mathcal{F}}^{G,q,(1)}  + 4\beta_0\hat{\mathcal{F}}^{G,q,(2)} + 2\hat{\mathcal{F}}^{G,q,(2)}\gamma^{(1)}_{GG} 
 \nonumber\\
 &+ 2\hat{\mathcal{F}}^{Q,q,(2)}\gamma^{(1)}_{GQ} + \hat{\mathcal{F}}^{G,q,(1)}\gamma^{(2)}_{GG} + \hat{\mathcal{F}}^{Q,q,(1)}\gamma^{(2)}_{GQ} + \frac{2}{3}\gamma^{(3)}_{GQ}  \Big\} + \hat{\mathcal{F}}^{G,q,(3)}
 \nonumber\\
 \mathcal{F}^{Q,g,(1)} =& \frac{2}{\epsilon}\gamma^{(1)}_{QG} + \hat{\mathcal{F}}^{Q,g,(1)}
 \nonumber\\
 \mathcal{F}^{Q,g,(2)} = &\frac{2}{\epsilon^2}\Big\{ \beta_0\gamma^{(1)}_{QG} + \gamma^{(1)}_{GG}\gamma^{(1)}_{QG} + \gamma^{(1)}_{QG}\gamma^{(1)}_{QQ} \Big\} + \frac{1}{\epsilon}\Big\{  2\beta_0\hat{\mathcal{F}}^{Q,g,(1)} +  2\hat{\mathcal{F}}^{G,g,(1)}\gamma^{(1)}_{QG} 
 \nonumber\\
 &+ 2\hat{\mathcal{F}}^{Q,g,(1)}\gamma^{(1)}_{QQ} + \gamma^{(2)}_{QG}    \Big\} + \hat{\mathcal{F}}^{Q,g,(2)}
 \nonumber\\
 \mathcal{F}^{Q,g,(3)} =&\frac{1}{\epsilon^3}\Big\{  \frac{8}{3}\beta_0^2\gamma^{(1)}_{QG} + 4\beta_0\gamma^{(1)}_{GG}\gamma^{(1)}_{QG} + \frac{4}{3}(\gamma^{(1)}_{GG})^2\gamma^{(1)}_{QG} + \frac{4}{3} \gamma^{(1)}_{GQ}(\gamma^{(1)}_{QG})^2  + 4\beta_0\gamma^{(1)}_{QG}\gamma^{(1)}_{QQ}  
 \nonumber\\
 &+\frac{4}{3}\gamma^{(1)}_{GG}\gamma^{(1)}_{QG}\gamma^{(1)}_{QQ} + \frac{4}{3}\gamma^{(1)}_{QG}(\gamma^{(2)}_{QQ})^2 \Big\} + \frac{1}{\epsilon^2}\Big\{ 4\beta_0^2\hat{\mathcal{F}}^{Q,g,(1)} + \frac{4}{3}\beta_1\gamma^{(1)}_{QG} 
 \nonumber\\
 &+ 6\beta_0\hat{\mathcal{F}}^{G,g,(1)}\gamma^{(1)}_{QG} +2\hat{\mathcal{F}}^{G,g,(1)}\gamma^{(1)}_{GG}\gamma^{(1)}_{QG} + 2\hat{\mathcal{F}}^{Q,g,(1)}\gamma^{(1)}_{GQ}\gamma^{(1)}_{QG} + 6\beta_0\hat{\mathcal{F}}^{Q,g,{1}}\gamma^{(1)}_{QQ} 
 \nonumber\\
 &+ 2\hat{\mathcal{F}}^{G,g,(1)}\gamma^{(1)}_{QG}\gamma^{(1)}_{QQ} +2\hat{\mathcal{F}}^{Q,g,(1)}(\gamma^{(1)}_{QQ})^2  + \frac{2}{3} \gamma^{(1)}_{QG}\gamma^{(2)}_{GG} + \frac{4}{3} \beta_0\gamma^{(2)}_{QG} + \frac{4}{3} \gamma^{(1)}_{GG}\gamma^{(2)}_{QG} 
 \nonumber\\
 &+ \frac{2}{3} \gamma^{(1)}_{QQ}\gamma^{(2)}_{QG}+\frac{4}{3}\gamma^{(1)}_{QG}\gamma^{(2)}_{QQ}  \Big\} + \frac{1}{\epsilon}\Big\{  \beta_1\hat{\mathcal{F}}^{Q,g,(1)} + 4\beta_0\hat{\mathcal{F}}^{Q,g,(2)} + 2\hat{\mathcal{F}}^{G,g,(2)}\gamma^{(1)}_{QG} 
 \nonumber\\
 &+ 2\hat{\mathcal{F}}^{Q,g,(2)}\gamma^{(1)}_{QQ}+ \hat{\mathcal{F}}^{G,g,(1)}\gamma^{(2)}_{QG} + \hat{\mathcal{F}}^{Q,g,(1)}\gamma^{(2)}_{QQ}  + \frac{2}{3}\gamma^{(3)}_{QG}  \Big\} + \hat{\mathcal{F}}^{Q,g,(3)}
 \nonumber\\
 \mathcal{F}^{Q, q,(1)}  = & \frac{2}{\epsilon} \gamma^{(1)}_{QQ} +  \hat{\mathcal{F}}^{Q,q,(1)} 
\nonumber\\
 \mathcal{F}^{Q,q, (2)}   = & \frac{2}{\epsilon^{2}} \Big\{\gamma^{(1)}_{GQ}\gamma^{(1)}_{QG} + \beta_{0}\gamma^{(1)}_{QQ} + (\gamma^{(1)}_{QQ})^{2} \Big\}
+ \frac{1}{\epsilon} \Big\{2\beta_{0}\hat{\mathcal{F}}^{Q,q,(1)} + 2\hat{\mathcal{F}}^{G,(1)}_{q}\gamma^{(1)}_{QG} 
 \nonumber\\&
 + 2\hat{\mathcal{F}}^{Q,q,(1)}\gamma^{(1)}_{QQ} + \gamma^{(2)}_{QQ}\Big\} + \hat{\mathcal{F}}^{Q,q,(2)}
 \nonumber\\
\mathcal{F}^{Q, q,(3)}  = & \frac{1}{\epsilon^{3}} \Big\{4\beta_{0}\gamma^{(1)}_{GQ}\gamma^{(1)}_{QG} + \frac{4}{3} \gamma^{(1)}_{GG}\gamma^{(1)}_{GQ}\gamma^{(1)}_{QG}
     + \frac{8}{3}\Big\{ \beta_{0}^2\gamma^{(1)}_{QQ} 
     + \gamma^{(1)}_{GQ}\gamma^{(1)}_{QG}\gamma^{(1)}_{QQ}\Big\}
 \nonumber\\&
 + 4 \beta_{0}(\gamma^{(1)}_{QQ})^2 + \frac{4}{3}(\gamma^{(1)}_{QQ})^3 \Big\}
       + \frac{1}{\epsilon^{2}} \Big\{4\beta_{0}^2\hat{\mathcal{F}}^{Q,q,(1)} + 6\beta_{0}\hat{\mathcal{F}}^{G,q,(1)}\gamma^{(1)}_{QG} 
       \nonumber\\&
       + 2\hat{\mathcal{F}}^{G,q,(1)}\gamma^{(1)}_{GG}\gamma^{(1)}_{QG}
 + 2\hat{\mathcal{F}}^{Q,q,(1)}\gamma^{(1)}_{GQ} \gamma^{(1)}_{QG}
 + \frac{4}{3}\beta_{1}\gamma^{(1)}_{QQ} +
     6\beta_{0}\hat{\mathcal{F}}^{Q,q,(1)}\gamma^{(1)}_{QQ}
  \nonumber\\&
  + 2\hat{\mathcal{F}}^{G,q,(1)}\gamma^{(1)}_{QG}\gamma^{(1)}_{QQ}
  + 2\hat{\mathcal{F}}^{Q,q,(1)}(\gamma^{(1)}_{QQ})^2 +
     \frac{2}{3}\gamma^{(1)}_{QG}\gamma^{(2)}_{GQ}
    + \frac{4}{3}\gamma^{(1)}_{GQ}\gamma^{(2)}_{QG}+ \frac{4}{3}\beta_{0}\gamma^{(2)}_{QQ}
    \nonumber\\&
    + 2\gamma^{(1)}_{QQ}\gamma^{(2)}_{QQ}\Big\}
     + \frac{1}{\epsilon}\Big\{\beta_{1}\hat{\mathcal{F}}^{Q,q,(1)}
       + 4 {\beta _{0}} \hat{\mathcal{F}}^{Q,q,(2)} + 2\hat{\mathcal{F}}^{G,q,(2)}\gamma^{(1)}_{QG}
  + 2\hat{\mathcal{F}}^{Q,q,(2)}\gamma^{(1)}_{QQ} 
  \nonumber\\&
  + \hat{\mathcal{F}}^{G,q,(1)}\gamma^{(2)}_{QG}
  + \hat{\mathcal{F}}^{Q,q,(1)}\gamma^{(2)}_{QQ} + \frac{2}{3} \gamma^{(3)}_{QQ} \Big\} + \hat{\mathcal{F}}^{Q,q,(3)}
 \end{align}
The explicit results of the above renormalised FFs can be obtained from the authors on request.

\subsection{Leading Transcendentality principle}
Recently, on-shell form factors in supersymmetric Yang-Mills theory have attracted a lot of attention
to understand their field theoretic structure. 
There are already several results~\cite{vanNeerven:1985ja, Brandhuber:2010ad, Gehrmann:2011xn} in ${\cal N}=4$ super Yang-Mills (SYM) 
with gauge group SU(N).
${\cal N}=4$ SYM is UV finite in $d=4$ dimensions and also dual to
type IIB string theory on $AdS_5 \times S^5$ with self dual RR field strength. 
This implies that one can relate quantities computed 
in ${\cal N}=4$ SYM in the strong coupling limit with those
obtained in the  weak coupling limit of the gravity theory. 
There have been efforts to compute on-shell amplitudes, correlation functions and form factors 
in SYM using perturbative approach to very good accuracy to make non-perturbative predictions
through systematic resummation procedures.  The important advances
in this direction include resummation of perturbative contributions~\cite{Anastasiou:2003kj,Bern:2005iz} to MHV amplitudes to all orders in 
`t Hooft coupling.  The developments in this direction have not only improved our understanding
of the quantum field theories in general but also provided very sophisticated analytical tools
to compute multi-loop multi-leg processes that are essential in the present collider phenomenology.         
Thanks to maximal supersymmetry in ${\cal N}=4$, large cancellations between various
contributions result in elegant and simple looking predictions that have lot of resemblance
with those in QCD.            
For example, the leading transcendentality principle~\cite{Kotikov:2004er, Kotikov:2006ts, Kotikov:2001sc}
relates anomalous dimensions of the twist two operators in
${\cal N} =4$ SYM to the leading transcendental (LT) terms of such 
operators computed in QCD.

Due to the presence of massless modes both in QCD and SYM,
the IR divergences show up when loop corrections are involved. 
The IR regulated results 
for the scattering amplitudes and form factors can be expressed as a
linear combinations of polylogarithmic functions whose maximum degree
of transcendentality depends on the order of perturbation theory.
Unlike QCD which receives contributions from all degrees of transcendentality up to $2l$, where '$l$' denotes the order in perturbative expansion, 
certain scattering amplitudes and FFs in $\mathcal{N} = 4$ SYM exhibit uniform transcendentality at each order.

Interesting relation between QCD quark and gluon
form factors~\cite{Gehrmann:2010ue}
and scalar form factor in SYM has been observed up to three loop.  If we  
replace \cite{Kotikov:2006ts} the colour factors $C_A = C_F = N$
and $ n_{f}=N$ in the quark and gluon form factors, then their
LT parts 
not only coincide with each other but also become identical, to the form factors of 
half-BPS scalar operator in ${\cal N}=4$ SYM
\cite{Gehrmann:2011xn}.
Similar behaviour was observed for the diagonal pseudo-scalar form factors ${\cal F}^{G,g}$ and
${\cal F}^{J,q}$ in \cite{Ahmed:2015qpa}.
A similar relation for three point form factors at two loop level between 
LT terms of $H\rightarrow ggg$ in QCD ~\cite{Gehrmann:2011aa}  and those of half-BPS operator in $\mathcal{N}=4$ SYM were found in ~\cite{Brandhuber:2012vm}.


In the present context, we have found the LT terms of 
the diagonal form factors, $\hat {\cal F}^{G,g},\hat {\cal F}^{Q,q}$ 
with the above prescribed colour replacement, are not only identical to
each other but also 
coincide,
with the LT terms of the scalar form factors in ${\cal N}=4$ SYM
\cite{Gehrmann:2011xn}.  This is true for terms proportional to positive
powers of $\epsilon$ available up to transcendentality 8 ~\cite{Lee:2010ik}.  
On the other hand, the LT terms of the off-diagonal ones namely, $\hat {\cal F}^{G,q}$, $\hat {\cal F}^{Q,g}$,
while identical
to each other after the replacement of colour factors, 
do not coincide with those of the diagonal ones.

\section{Conclusions}

In this article, we have studied in detail the theoretical issues with 
the interactions of spin-2 fields with those of the SM.  We
have considered a set of gauge invariant tensorial operators
constructed out of fields of the SM that couple to spin-2 fields.  These
operators are in general not conserved like the usual EM 
tensor.  Hence they require additional renormalisation.  To compute these
additional renormalisation constant, we have exploited the universal
infrared structure of on-shell amplitudes with composite operators.
Computing these form factors order by order in perturbation theory and
using the K-G equation we obtain the UV anomalous dimensions and the
renormalisation constants up to three loop level.  The renormalisation
constants and the on-shell FFs are important components of
observables that can probe the physics of spin-2 fields.  We have
reserved the detailed phenomenological study with these two operators
at the LHC for future publication \cite{Ahmed:2016}.

\begin{acknowledgments}
We would like to thank T. Gehrmann, R. Lee and F. Maltoni for useful discussions and timely help.
\end{acknowledgments}

\bibliographystyle{JHEP}
\bibliography{main}

\end{document}